\documentstyle{article}
\newcommand{\beq}{\begin{equation}}
\newcommand{\eeq}{\end{equation}}
\newcommand{\bea}{\begin{eqnarray}}
\newcommand{\eea}{\end{eqnarray}}
\newcommand{\q}{\mbox{\boldmath $q$}}
\newcommand{\p}{\mbox{\boldmath $p$}}
\newcommand{\Q}{\mbox{\boldmath $Q$}}
\newcommand{\bP}{\mbox{\boldmath $P$}}
\newcommand{\A}{{\bf A}}
\newcommand{\bS}{{\bf S}}
\newcommand{\cn}{\mbox{cn}}
\newcommand{\sn}{\mbox{sn}}
\begin{document}

\title{Exact solutions in the FPU oscillator chain}

\author{P. Poggi$^1$ \hspace{0.5in} S. Ruffo$^{1,2}$\thanks{e-mail:
ruffo@ingfi1.ing.unifi.it}  \\
$^1$ {\small\it Dipartimento di Energetica, Universit\`{a} di Firenze,}\\
{\small\it via S. Marta 3, I-50139 Firenze (Italy).} \\
$^2$ {\small\it INFN and INFM, Firenze (Italy).}}

\date{16 October 1995}

\maketitle

\begin{abstract}
After a brief comprehensive review of old and new results on the well known
Fermi-Pasta-Ulam (FPU) conservative system of $N$ nonlinearly coupled
oscillators, we present a compact linear mode representation of the
Hamiltonian of the FPU system with quartic nonlinearity and periodic
boundary conditions, with explicitly computed mode coupling coefficients.
The core of the paper is the proof of the existence of one-mode and two-mode
exact solutions, physically representing nonlinear standing and travelling
waves of small wavelength whose explicit lattice representations are
obtained, and which are valid also as $N\rightarrow \infty$. Moreover, and
more generally, we show the presence of multi-mode invariant submanifolds.
Destabilization of these solutions by a parametric perturbation mechanism
leads to the establishment of
chaotic in time mode interaction channels, corresponding to the formation
in phase space of bounded stochastic layers on submanifolds.
The full mode-space stability problem of the $N/2$ zone-boundary mode is
solved, showing that this mode becomes unstable through a mechanism of the
modulational Benjamin-Feir type. In the thermodynamic limit the mode
is always unstable
but with instability growth rate linearly vanishing with energy density.
The physical significance of these solutions and of their stability properties,
with respect to the previously much more studied equipartition problem
for long wavelength initial excitations, is briefly discussed.
\end{abstract}


\section{Introduction}
\label{Introduction}

The numerical experiment by Fermi, Pasta, Ulam (FPU)
and Tsingou\footnote{M. Tsingou
contributed to the numerical work and then did not participate
in the writing of the report.}~\cite{Fput},
perhaps the first of all times,
was the first attempt to check the predictions of classical statistical
mechanics concerning the dynamics of a Hamiltonian system
of coupled oscillators with a large number $N$ of degrees of freedom.
The result of this experiment was a big
surprise: the expected relaxation to equipartition of energy among the
linear normal modes was not revealed, during the time of observation
and with low energy initial excitations.
This also implies that ergodicity is not an obvious consequence of the
non-existence of analytic first integrals of the motion besides
the total energy (and possibly total momentum).
After an initial growth of the energy in the neighbouring modes
(as expected) they observed that the energy sharing was restricted only to
the first modes, with a quite regular dynamics, rather than detecting
a gradual and continuous energy flow from the first excited mode
to the higher ones and a stochastic dynamics.
Even more surprisingly, at later times, almost all the energy was
flowing back into the initially excited mode, so that the system seemed
to possess quasiperiodicity properties.
Chirikov et al.~\cite{Izrailev66} showed later,
also with numerical experiments~\cite{Chir-Izr-Tayursky73},
that, at sufficiently high energies, the FPU model did relax to the
equipartition state, on times which become smaller and smaller
as the energy is increased.
It then became clear that such a system
was displaying qualitatively different behaviours as the energy,
fixed by the initial condition, was varied.
Chirikov and Izrailev~\cite{Izrailev66} gave also the first
linear normal modes representation of the system and applied
the nonlinear resonances theory to predict the existence of the
transition between the two behaviours.
They did also recognize the resemblance between these results
and the Kolmogorov-Arnol'd-Moser~\cite{KAM} theorem, first formulated,
although not completely demonstrated, by Kolmogorov
just in the same year of the FPU numerical experiment.
These results successively stimulated a lot of
numerical studies aiming at determining the dependence of the different
observed behaviours of the FPU system on the number $N$ of degrees
of freedom (see Ref.~\cite{Ben-Varenna} and references cited therein,
and also~\cite{Livi85,Kantz89,Pettini90,Galgani92}).

It is now well established that the transition between a quasi-integrable
behaviour to a mixing one, in the FPU system as well as
in similar models  (Lennard-Jones in 1-D and 2-D, $\phi^4$),
is controlled by the energy.
At small energies the motion is weakly chaotic with positive but small Lyapunov
exponents, revealing the presence of thin stochastic layers in the phase
space, which is mostly filled with KAM tori.
At higher energies the maximum Lyapunov exponent and the
Kolmogorov-Sinai entropy rise considerably, revealing the growth
of stochastic regions. In this energy range the temporal signals
of normal mode energies also become chaotic.
Well above the transition region all the signatures of large-scale chaos
are present, with $N-1$ ($N-2$ for systems with periodic boundary conditions)
positive Lyapunov exponents, fast diffusion of the orbit over the
constant energy surface, rapidly decaying spatio-temporal
correlation functions (also between the modal energies).
It has been also shown~\cite{Livi86} that, in this region,
a density function of Lyapunov exponents can be introduced
in the thermodynamic limit, which can also be analytically estimated with
a random matrix approximation~\cite{Eckmann}.

On the other hand, the quasiperiodic behaviour at low energy
was interpreted in terms of a continuum limit of the model.
It was in fact observed by Zabusky and Kruskal~\cite{Zabusky65}
that, increasing the number $N$ of oscillators while holding
constant the length of the chain, the FPU model with cubic
nonlinearity can be reduced to the Korteweg-deVries (KdV) equation
in the limit of low amplitude and long wavelength excitations,
after selecting one of the two possible directions of propagation
for travelling waves.
For the KdV equation the same authors found numerically the existence of
localized solutions which preserve their shape even after mutual collisions.
To emphasize the extremely stable and localized character of such
solutions Zabusky and Kruskal called them {\em solitons}.
It was shown later~\cite{Gardner67, Lax68, Zakharov71} that the KdV
equation is a completely integrable infinite-dimensional Hamiltonian system,
whose initial value problem can be solved by the
spectral tranform method~\cite{AblowitzKNS73,Ablowitz-books}.
The quasiperiodic behaviour of the FPU system at low energies
is consequently believed to be the result of the presence on the
lattice of several solitons emerging from the long wavelength initial condition
and travelling with different speeds so that, at particular times,
they reproduce almost exactly the initial condition.
Quasiperiodic behaviour is destroyed at larger energies, above
the threshold of transition to equipartition~\cite{Livi85}.
A deeper understanding of this transition phenomenon requires a systematic
study of its dependence on the number of degrees of freedom,
on the initial condition and on the time of observation.
Such a study has not been fully completed yet.

\subsection{The FPU model}
\label{FPU-model}

The FPU system is an approximate model for the behaviour of a classical solid
at low temperatures. The reduction of complexity in comparison to the
real physical situation is considerable. First of all only one spatial
dimension is considered and then the interaction (typically of the
Lennard-Jones kind) is expanded in the small displacements
around the equilibrium positions of the atoms or molecules,
i.e. the weakly anharmonic case is considered (in practice the case of
rather low temperatures).
As a consequence, several problems must be discussed concerning
the physical relevance of the FPU model. We must first
ask if the observed phenomena are peculiar of the model or can be
instead revealed in a wider class of more physical models.
This issue is now clarified and we can state with no doubt that different
models in one and two dimensions share the same phenomenology
(see Ref.~\cite{Ben-Varenna} for a review). In particular,
practically all the models which, as the FPU one, are constituted
by weakly coupled anharmonic oscillators, show the prevalence of
ordered motion at sufficiently low energies, while at higher energies
their phase space becomes chaotic.
An important point is that the transition between the
``ordered'' region and the ``chaotic'' one happens in an energy range
relevant for physics. For example, in a model of oscillators
in two dimensions with Lennard-Jones interactions~\cite{Ben-Ten},
the transition, with the physical parameters of Argon,
is found to happen around a temperature of $5~^{\circ}K$.
Some recent molecular dynamics calculations~\cite{CLMP95} on a
model for a Xenon crystal with diluted impurities made of Iodine molecules,
show that the time required for the equipartition of energy
between the initially excited $\mbox{I}_2$ molecule and the Xe matrix
abruptly increases at approximately $30~^{\circ}K$;
such a physical situation could be investigated in a real
experimental sample.

Having established the physical relevance of the phenomena studied
in the FPU model, we proceed to its definition.
The FPU model is a one-dimensional chain of oscillators
with unit mass, with weakly nonlinear nearest-neighbour interaction
(the lattice spacing is also taken of unitary length).
Calling $q_i$ and $p_i$ the coordinates and, respectively, the momenta
of the oscillators, the model is defined by the following Hamiltonian:
\beq
H=\sum_{i=1}^N \frac{p_i^2}{2} + \sum_{i=1}^N
\left[ \frac{1}{2} (q_{i+1} -q_i)^2
+\frac{\beta}{r}(q_{i+1} -q_i)^r \right]~,
\label{hamilton-modello}
\eeq
where $q_1 \equiv q_{N+1}$ and $r=3$, for the so-called FPU-$\alpha$ model,
while $r=4$ for the FPU-$\beta$ model. We consider in the present work only
the FPU-$\beta$ model, with periodic boundary conditions to allow
travelling wave solutions (whereas in the original FPU paper fixed ends
boundary conditions were chosen).
At fixed energy, the coupling constant $\beta$ determines the
amount of nonlinearity in the model. Conversely, as it is more natural,
for a fixed $\beta$ the increasing departure from the harmonic behaviour
is controlled increasing the energy. It can be easily shown that the
dynamics depend only on the parameter $\beta E^{(r/2 -1)}$,
where $E$ is the total energy, fixed by the initial condition.

Hamiltonian~(\ref{hamilton-modello}), written in linear normal coordinates
$(Q_k,P_k)$ (phonons) becomes
\beq
H= \frac{1}{2} \sum_k \left( P_k^2 + \omega_k^2 Q_k^2 \right)
+\beta V(\Q)~.
\label{hamilton-normale}
\eeq
with frequencies $\omega_k = 2 \sin(\pi k/N)$
in the case of periodic boundary conditions
(the harmonic frequency spectrum is obvioulsy different from the case
of fixed boundary conditions studied by FPU) and we have
$\omega_k=\omega_{N-k}$, so that there are only $N/2$ different
frequencies (if $N$ is even, for simplicity).
The harmonic energy of mode $k$ is defined by
$E_k= (P_k^2 +\omega_k^2 Q_k^2)/2$.

The FPU experiment aimed at showing the progressive decorrelation of
the system during its temporal evolution, eventually leading to
ergodic behaviour. To this end they chose an initial condition far from
equilibrium, giving all energy to the lowest ($k=1$) normal mode only,
and then calculating the instantaneous energies $E_k(t)$ of all modes.
They expected to see a progressively uniform redistribution of energy
between all modes, caused by the small anharmonic coupling between them.
On the contrary they observed the well known FPU recurrent behaviour:
energy was flowing back regularly to mode $k=1$ after an initial share.
Return to the initial condition is not exact, but the possibility that
relaxation is present on longer times was ruled out by the numerical
experiment of Tuck and Menzel~\cite{Tuck}, who first observed the
``superperiod'' phenomenon.

The behaviour of average energies $\bar{E}_k(T) =T^{-1} \int_0^T E_k(t) dt$
versus $T$, shows that the systems relaxes to an asympotic state,
but the latter is different from that in which equipartition of energy holds.
On the contrary, at higher energy~\cite{Chir-Izr-Tayursky73},
the equipartition state is reached in a relatively short time.
A transition is present from an ``ordered''
state, in which the system seems not to be relaxing towards equipartition,
showing  recurrent behaviour in time, to a different one where, on the
contrary,
equipartition is quickly reached, which we call a ``chaotic'' state
(though these terms are improper because also in the ``ordered''
state there are chaotic solution, but they fill a part of small measure
in the phase space).

Various indicators have been employed
to characterize this transition, either based on the distribution
of energy among the normal modes~\cite{Kantz89},
or on the local divergence rate
of nearby trajectories in phase space (Lyapunov exponents)~\cite{Ben-Ten}.
In particular, the first clear numerical evidence of the transition
to equipartition (or ``chaotic'') state was obtained in
Ref.~\cite{Livi85} by the use of an appropriate Shannon entropy $S$ defined
in the space of modes (``spectral entropy'').
At a given value of the control parameter
\beq
\epsilon=\frac{\beta E}{N}~,
\label{epsilon}
\eeq
where $E$ is the total energy, the ``spectral entropy'' was shown
to increase, reaching then its maximal value.

Despite these results the problem is still far from beeing understood,
especially concerning the dependence of the phenomenology
on the number of degrees of freedom and on the time scales
of observation.

\subsection{Recent results}
\label{Recent-results}

We will focus on the results which are valid at large $N$
(thermodynamic limit).
The first result we want to mention concerns the Lyapunov spectrum
well above the region of transition to equipartition ($\epsilon \gg 1$).
There are strong numerical evidences~\cite{Livi86} supporting the
possibility to define a function $\phi(x)$ which represents the density
of Lyapunov exponents in the thermodynamic limit,
in the case when the only vanishing exponents are those related
to space and time translational symmetries (momentum and energy
conservation, respectively). After labelling the Lyapunov exponents
in the decreasing order $\lambda_1 \geq \lambda_2 \geq \dots \geq
\lambda_{2N}$,
the density function $\phi(x)$ is given by
\beq
\phi(x) = \lim_{N \to \infty} \lambda_{xN} (N)~.
\label{distribuzione}
\eeq
The existence of this limit was conjectured by Ruelle~\cite{Ruelle82}
for the Navier-Stokes equation, and verified by Manneville~\cite{Manneville}
for the Kuramoto-Sivashinsky equation.
The existence of such a function makes it possible to define
(using the Pesin relation) the Kolmogorov-Sinai entropy
as an intensive quantity in the thermodynamic limit,
thus proving the persistence of a state of large-scale chaos
also when the number of degrees of freedom goes to infinity.
So, when the intensive control parameter $\epsilon$ is big enough,
the trajectory of the system visits in a short time all accessible
regions of phase space and, as a consequence,
equipartition of energy among the normal modes is quickly reached.
The evidence for the existence of the Lyapunov exponents density function
shows, in an indirect way, that equipartition of energy in the FPU
model occurs also in the thermodynamic limit, when the energy
density is high enough. Unfortunately there are no analytical
results proving the existence of the function $\phi(x)$,
apart from those obtained by Eckmann and Wayne using a
random matrix approximation~\cite{Eckmann}.
A recent discussion of this
problem was also given by Sinai~\cite{Sinai}, who has indeed obtained a
rigorous proof of the existence of an analogous quantity,
obtained interchaning the order of the limits $N \rightarrow \infty$
and $t \rightarrow \infty$ (the latter appearing in the definition
of the Lyapunov exponents). Though this is not, a priori, the same thing as
$\phi$, it can be argued~\cite{Sinai} to be more relevant for statistical
mechanics; moreover it has a closer correspondence to the
numerical procedures used to determine $\phi$.

Pettini and Landolfi~\cite{Pettini90} found a rapid decrease of the
maximum Lyapunov exponent $\lambda_1$ as the control parameter $\epsilon$
is decreased below a value $\epsilon \approx 0.1$; such a behaviour
does not change significantly when $N$ is increased.
Even if this does not necessarily prove the onset of large-scale chaos
(because only a few directions, among many, might be responsible for
the increase of the divergence rate), the persistence
of the result when $N$ is varied reveals the presence of an important
phenomenon occurring in phase space.
This result is consistent with the results obtained using the
``spectral entropy'', which also sharply increases around
$\epsilon \approx 0.1$ (although a slower diffusion of the entropy with time
is also present for smaller values of $\epsilon$)~\cite{DeLuca95,KLR94,DLR95}.

More recently~\cite{Pettini93} it has been suggested that
the largest Lyapunov exponent can be estimated from the temporal evolution
of the vector \mbox{\boldmath $\xi$} representing the deviation between
geodesics
on the Riemannian manifold constructed with the Eisenhart metric
over the enlarged configuration space-time. In particular, it was shown that
the norm $\psi = \| \mbox{\boldmath $\xi$} \|$ obeys the equation
\beq
\ddot{\psi} = - K_R \psi
\label{schroe}
\eeq
where $K_R = \nabla^2 V(\q)/N$ is the positive Ricci curvature of the manifold
and $V$ the potential of the FPU model.
Supposing that the quantity $K_R$ and the magnitude of its fluctuations
can be calculated, in the thermodynamic limit, using the Gibbs equilibrium
measure and solving for the instability growth rate of Eq.~(\ref{schroe}),
Casetti, Livi and Pettini~\cite{Pettini95} obtain values of the maximum
Lyapunov exponent compatible with those reported in~\cite{Pettini90}.
The exponent would be zero in the absence of temporal fluctuations of $K_R$.
It is an intriguing fact that the time-periodic dependence of $K_R$ on
stable periodic (elliptic) orbits must differ enough from
that on unstable periodic
(hyperbolic) orbits, even if they are close in phase space,
to make the Lyapunov exponent vanish. In fact, the positivity of
Lyapunov exponent is caused, in this case, by the fact of being out of
the spectrum of the periodic operator in~(\ref{schroe}).
A more complicated case is the one of ``chaotic'' functions $K_R$.
It is treated in Ref.~\cite{Pettini95} and has a close resemblance
to the Anderson localization problem in one dimension.

Concerning perturbation theory results, Galgani, Giorgilli and collaborators
in a recent paper~\cite{Galgani92} summarize the findings obtained
using Nekhoroshev estimates~\cite{Nekho}. The latter permit to
evaluate upper bounds for the time variation of the unperturbed actions
on times that, though being finite, increase exponentially
as the perturbation parameter is decreased
(in the FPU case the $\beta$ parameter of the nonlinear term).
It is possible, using this approach, to find results valid
for initial conditions on open sets in the phase space, as opposed
to methods based on the Kolmogorov-Arnold-Moser theorem
(on the other hand the latter has the advantage to give statements
valid for all times).
The stability time $\tau$ of the single unperturbed actions
(or action ``freezing'' time) is found to scale as
\beq
\tau = \tau_* \exp (\frac{\beta_*}{\beta})^d
\label{tempo}
\eeq
where, in general, both $\tau_*$, $\beta_*$ and $d$ depend on $N$.
The most important dependence on $N$ is that of $d$: the best
estimated so far obtained for FPU gives $d \simeq N^{-1}$,
a result confirmed also by numerical simulations
(so the estimate seems to be optimal). This result suggests that
in the thermodynamic limit the freezing times might become short, or even
vanishing, and the region of violation of energy equipartition could disappear.
We must, however, remember that such estimates are valid in an energy region
shrinking to zero as $N$ is increased, and moreover they are lower bounds,
so they could be irrelevant.
A positive result was however obtained on a modified
FPU model with alternating light and heavy masses (like in a diatomic solid).
In this case the harmonic frequency spectrum splits in two separated
components: an ``acoustic branch'' with low frequencies and an
``optical branch'' at high frequencies; the latter is almost completely
degenerate as the ratio between the masses is increased and $N \to \infty$.
Then it is possible to give Nekhoroshev estimates for subsets of unperturbed
actions as the total harmonic energy of the optical modes,
which turns out to be frozen  on a time given by the law~(\ref{tempo}),
but now with $d=1$, $\tau_* \simeq  N^{-6}$ and $\beta_* \simeq N^{-8}$.
Such analytical estimates suggest again the vanishing of the freezing time
in the thermodynamic limit, but in this case numerical simulations
show that the estimates are far from being optimal:
 $\beta_*$ seems to be independent of $N$, while $\tau_*$ has a weak
dependence on $N$, possibly like $(\ln N)^{-1}$. The authors of
Ref.~\cite{Galgani92}
then argue in support of the relevance of Nekhoroshev's like estimates
in the thermodynamic limit.
Unfortunately, for the FPU model it is not easy to identify a natural
decomposition in subsets of unperturbed actions. Perhaps this might
be possible in the light of the present work, where we show the existence
of subsets of normal modes where energy remains trapped for suitable
initial conditions, so defining reduced (integrable or not) Hamiltonians
on submanifolds of the constant energy hypersurface.
It would be interesting to attempt Nekhoroshev-like estimates
when the energy is restricted to these submanifolds.

More recently De Luca, Lichtenberg and Lieberman~\cite{DeLuca95} studied
the approach to equipartition in the FPU model exploiting normal form theory
to find an effective Hamiltonian describing the interaction among a
reduced number of modes (four in their treatment).
The main result of their theory is that above a critical energy $E_c$
the system reaches a near-equipartition state, in a time proportional to $N^2$;
below this critical energy the time needed increases even faster
with $N$ (perhaps exponentially).
This holds when the initial excitation is given to a subset of low
modes whose number does not increase with $N$.
If instead the excited modes are a subset scaling proportionally to $N$
($k \propto N)$, the typical time scale to quasi-equipartition increases
like $N$. These predictions are also supported by numerical simulations.
The idea to construct the reduced Hamiltonian is to make a large $N$
expansion of the dispertion relation
$\omega_k = 2 \sin [\pi k/2(N+1)] \simeq [\pi k /(N+1)] + O((k/N)^3)$
(we are treating now the system with fixed ends)
and to identify the four-wave resonance relations
$(k_1 + k_2 + k_3 + k_4 =0)$
producing in the resonant normal form those angles which are slowly
(adiabatically) varying; these latter are found to be
$\theta_s =\theta_1 +\theta_3 -2\theta_2$ and
$\theta_{sp}  =\theta_2 +\theta_4 -2\theta_3$.
The parameter controlling the deformation of the actions
(monotonic in the energy) is seen to be proportional to $NE$,
while the angles $\theta_s$ and $\theta_{sp}$ are slowly evolving
with the frequency $\Omega \sim kE/N^2$; the latter determines the
characteristic evolution time for the actions $\tau \sim N^2/E$ for
$k \simeq \mbox{const.}$, while $\tau \sim N/E$ if $k \propto N$.
The energy transfer to higher modes is present,
but takes place on much longer times.
Actually, it is also known that the energy fraction transferred to
the highest modes is exponentially small in
mode number~\cite{Parisi,Livi83-exponential}.
De Luca et al.'s idea is to use resonant normal forms and expand the dispersion
relation for large $N$ to obtain an effective Hamiltonian.
This latter expansion was also suggested by Shepelyansky~\cite{Shepelyansky},
however his purpose was to estimate the largest Lyapunov exponent which he was
able to show, by his method, to be positive even at very small energies.
So, according to Shepelyansky, the critical
energy for the onset of chaos goes to zero as $N$ is increased
(in agreement with what is obtained in~\cite{DeLuca95}).
This last result is consistent with the numerical result by Pettini and
Landolfi~\cite{Pettini90}, but the scaling law suggested by Shepelyansky
is not consistent with numerical results.

It is still difficult to arrange all these results in a single
coherent framework, and especially it is not clear if it can be really
justified to neglect the interaction with high modes.
In fact, we show in this paper that the structure of the interaction among
the modes is rather complex.

Kantz, Livi and Ruffo~\cite{KLR94}, indipendently from De Luca et al.,
revealed with numerical experiments that the typical time scale for
equipartition increases with $N$. More precisely, considering the
functional dependence of the ``spectral entropy'' $S$
on energy $E$, time $t$ and on the number $N$ of oscillators,
these authors found that $S$ depends only on the two variables $E$ and $t/N$:
$S (E,t,N)=f(E,t/N)$, for initial conditions with $k \propto N$.
Moreover, for initial conditions with $k \simeq \mbox{const.}$ they found
$S(E,t,N) = g(E/N,t/N)$.
This last result is in apparent contradiction with De Luca et al.,
who rather seem to suggest $ S(E,t,N) = g(E,t/N^2)$;
but if both results have to be valid, it is clear that $S$ must be
a function of the single parameter $Et/N^2$. This prediction has been
recently numerically confirmed~\cite{DLR95}.
Moreover, a completely independent analytical method based on the derivation
of the breakdown (shock) time for the non-dispersive limit of the mKdV
equation (to which the FPU-$\beta$ model reduces in the continuum limit)
leads to the same prediction~\cite{PRK95}. However, a correction to this
scaling theory due to a phase-space filling diffusive process is also
numerically revealed~\cite{DLR95},
which finally gives a dependence of $S$ on the
single parameter $\beta E k t/(N^2 \sqrt{N})$. It should be observed that even
in the thermodynamic limit at fixed $E/N$ and with $k \propto N$, will the
time scale of the system then diverge as $\sqrt{N}$. This is very intriguing
because it suggests (also in the spirit of Ref.~\cite{Sinai}) that
far from equipartition states might persist in time if we perform
the $N \rightarrow \infty$ limit before the $t \rightarrow \infty$
limit (a quite uncommon procedure for equilibrium statistical mechanics,
but perhaps a meaningful one in non-equilibrium).

The purpose of this paper is twofold. First, we want to show several
exact periodic and quasiperiodic solutions for the FPU-$\beta$ system
with periodic boundary conditions, whose majority appears to be new.
Those which have already appeared in the literature are rederived
by a unified method, which consists essentially in ``extending''
some harmonic modes to the anharmonic case.
Second, having at our disposal explicitly known solutions, we take
a first step in the study of their stability and its connection
with the onset of chaos in the system. To achieve these results
we rely on the structure of modal interactions in the FPU-$\beta$
model, which is derived in Section~\ref{Hamilton-normal}.
In Section~\ref{Integrable} we discuss the specific fully integrable case of
the three particle FPU-$\beta$ model. In Section~\ref{Explicit-solutions}
we derive explicit one-mode and two-mode solutions.
In Section~\ref{Stability} we identify two-mode and three-mode invariant
submanifolds where the motion displays
transitions to chaos after successive bifurcations.
Section~\ref{Stability} is moreover devoted to a
study of the restricted
stability in a two-mode submanifold and to the full phase-space stability
analysis of the $k=N/2$ zone-boundary solution.

\section{Linear normal coordinates \protect\\ for the FPU-$\beta$ system}
\label{Hamilton-normal}

Hamiltonian~(\ref{hamilton-modello}) of the FPU-$\beta$ system with
$N$ particles and periodic boundary conditions can be splitted in two
contributions:
\beq
H=H_{0}+H_{1}
\label{H}
\eeq
with
\bea
H_{0} & = & \frac{1}{2} \sum_{n=1}^{N} p_{n}^{2} +
\frac{1}{2} \sum_{n=1}^{N} (q_{n+1} - q_{n})^{2}~,  \label{H0}  \\
H_{1} & = & \frac{\beta}{4} \sum_{n=1}^{N} (q_{n+1} - q_{n})^{4}~,
\label{H1(q)}
\eea
where $q_{N+1} \equiv q_{1}$ and $\beta > 0$.

While the interaction in physical space has the simple structure
of nearest-neighbour coupling, it is clear that it becomes highly tangled
when it is represented in modal (i.e. Fourier) space.
Explicit expressions for the equations of motion in normal coordinates
appeared sporadically in the literature for the FPU-$\beta$ system in the
fixed boundary case, and they were used for varied purposes.
In Ref.~\cite{Izrailev66} the modal equations were treated with the
Krylov-Bogoliubov-Mitropolsky averaging technique, and then subjected to the
criterion of overlapping resonances~\cite{Chirikov79}
to obtain estimates for the threshold ofstochasticity.
Recently, in~\cite{DeLuca95,DLR95} the modal equations are the starting
point to develop approximations, based on normal form theory, to study
the time scales to equipartition in
the case when the energy is initially in a single or small group of
low-frequency modes.
Estimates for the FPU recurrence period were obtained by perturbative
treatment of the modal equations~\cite{Ford}.
Recently, Sholl and Henry~\cite{Sholl-Henry} applied
a shifted frequency perturbation scheme to the modal equations to
perform calculations of {\em superperiod} recurrence times,
at low anharmonicity, in FPU-$\alpha$ and FPU-$\beta$ chains with fixed ends.
In Ref.~\cite{Henry-Grindlay} the case of the FPU-$\beta$ chain
with fixed ends and 15 moving particles excited from
rest in the 11th mode was considered
(this is the most studied case were the so-called
{\em induction period} phenomenon~\cite{Saito,Bivins} shows up);
a detailed description of the early phases of the dynamics in the chain,
resulting from numerical integration, was given
and interpreted by means of the structure of modal couplings.

In our treatment we need to use some selection rules arising from the
structure of the equations of motion in Fourier space.
The first description of selection rules for the FPU-$\beta$ chain with
fixed ends was given by Bivins, Metropolis and Pasta~\cite{Bivins}
and later in more detail and for more general interaction potentials
by Sholl~\cite{Sholl,Sholl-tesi}. The case of periodic boundary conditions
is slightly more involved, due to the degeneracy of the harmonic frequency
spectrum, and to our knowledge it has been given only a very brief treatment,
with some missing points,
in unpublished work by Sholl~\cite{Sholl-tesi}.
It is not however our purpose to give here a complete description of
the selection rules arising in the periodic boundary case,
which could be given along the lines of Refs.~\cite{Sholl,Sholl-tesi},
and we will introduce only what is strictly necessary to derive our results.

\subsection{Normal coordinates for the periodic chain}
\label{Normal-coordinates}

Although the transformation to normal coordinates of the harmonic
Hamiltonian $H_{0}$ is quite an elementary topic, it is useful to
consider it in some detail to fix the notations and make some remarks.

The ``unperturbed''Hamiltonian $H_{0}$  can be written,
introducing the column vectors \q\ and \p\ of coordinates
and conjugate momenta, as
\begin{equation}
H_0 = \frac{1}{2} ~^{t}\p\p + \frac{1}{2} ~^{t}\q \A \q
\end{equation}
where $~^t \cdot$ denotes transposition and
$\A$ is the symmetric $N \times N$ matrix with elements
\begin{equation}
A_{i j}=-\delta_{i-1,j}+2\delta_{i,j}-\delta_{i+1,j}-
\delta_{i,1}\delta_{j,N}-\delta_{i,N}\delta_{j,1}~,~~~~
\begin{array}{l} i=1,\ldots,N \\
j=1,\ldots,N \end{array} \eeq
Let $\bS$ be a $N \times N$ real orthogonal ($\,^{t}\bS \bS={\bf I}$)
matrix which makes $\A$ diagonal through the similarity transformation
$\A \mapsto ~^{t}\bS\A\bS$.
Then we can perform a canonical transformation  from the old
set of coordinates
\[ (~^t\q \,;~^t\p)=(q_1,\ldots,q_N;p_1,\ldots,p_N) \]
to the new set of coordinates
\[ (~^t\Q \,;~^t\bP)=(Q_0,\ldots,Q_{N-1};P_0,\ldots,P_{N-1})~, \]
by the generating function
\beq
F(\q,\bP)=
\sum_{k=0}^{N-1} \sum_{n=1}^{N} S_{nk}P_k q_n~.
\label{generating}
\eeq
We have shifted the subscript of the new coordinates for reasons of
convenience, i.e. to label the center of mass motion by the zero index.
Through this transformation the harmonic part $H_{0}$ takes the form
of the Hamiltonian of a system composed of $N-1$ uncoupled harmonic
oscillators (the harmonic normal modes) plus a free particle
(representing center of mass motion).

In view of the application of this transformation to the full anharmonic
system it is important to remark that in the present case of periodic
boundary conditions, owing to the degeneracy of the
spectrum of $\A$, the matrix $\bS$ can be chosen in infinite
different ways, leading to different sets of normal coordinates which,
at variance with the harmonic case, are not equivalent
for the full anharmonic Hamiltonian. Therefore the set of normal coordinates
must be carefully specified.

Defining, for integer $k$,
\beq
\omega_{k}=2 \sin \left( \frac{\pi k}{N} \right)~,
\label{omega}
\eeq
the eigenvalues of $\A$ are $\mu_{k}=\omega_{k}^2$ for
$k \in \{0,1,\ldots,[N/2]\}$\footnote{We denote by $[x]$
the integer part of $x$, so that $ [N/2]= N/2 $ if $N$ is even
while $[N/2]=(N-1)/2 $ if $N$ is odd.},
and they have multiplicity two, apart from
$\mu_{0}$ (=0) and
(for even $N$ only\footnote{To keep generality while avoiding
frequent distinctions, from now on we keep the convention that
every proposition where the value $k=N/2$ appears is intended to be valid
only for the even $N$ case, while it can be simply omitted
for odd $N$. This is true, in particular for~(\ref{u^(N/2)})
and~(\ref{w}). } )
$\mu_{N/2}$ which are non-degenerate.

The uniquely defined (apart from a sign) normalized eigenvectors
${\bf u}^{(0)}$ and ${\bf u}^{(N/2)}$
corresponding respectively to $\mu_{0}$ and
$\mu_{N/2}$ have components
\begin{eqnarray}
u^{(0)}_{n}   & = & \frac{1}{\sqrt{N}}~,~~~~n=1,\ldots,N \\
u^{(N/2)}_{n} & = & \frac{(-1)^{n}}{\sqrt{N}}~,~~~~n=1,\ldots,N \label{u^(N/2)}
\end{eqnarray}
Because of the degeneracy of the spectrum, there is freedom in the
choice of an orthonormal basis in each two-dimensional eigenspace.
Defining for $k=1,\ldots,[(N-1)/2]$ the vectors ${\bf u}^{(k)}(\gamma)$,
containing the arbitrary real parameter $\gamma$, as
\beq
u^{(k)}_{n}(\gamma) =
\sqrt{\frac{2}{N}} \sin \left( \frac{2\pi kn}{N}+\gamma
\right)~,~~~~n=1,\ldots,N
\label{u}
\eeq
it is easy to verify that
\beq
\A{\bf u}^{(k)}(\gamma)=\mu_{k} {\bf u}^{(k)}(\gamma)~,~~~~\forall
\gamma \in \mbox{\boldmath $R$}
\eeq
and that the set of $N$ vectors
$\{{\bf w}^{(k)}(\gamma)\}_{k\in\{0,\ldots,N-1\}}$ given by
\beq
{\bf w}^{(k)}(\gamma)=\left\{ \begin{array}{ll}
{\bf u}^{(0)}         & \mbox{  for $k=0$} \\
{\bf u}^{(k)}(\gamma) & \mbox{  for $k=1,\ldots,[\frac{N-1}{2}]$} \\
{\bf u}^{(N/2)}       & \mbox{  for $k=N/2$} \\
{\bf u}^{(N-k)}(\gamma+\frac{\pi}{2}) &
\mbox{  for $k=[\frac{N}{2}]+1,\ldots,N-1$}
\end{array} \right.
\label{w}
\eeq
is an orthonormal basis composed by eigenvectors of $\A$
(for each arbitrarily fixed $\gamma$).
The eigenvectors corresponding to the eigenvalue
$\mu_{k}$ ($k=1,\ldots,[(N-1)/2]$) are
${\bf w}^{(k)}(\gamma) \equiv {\bf u}^{(k)}(\gamma)$ and
${\bf w}^{(N-k)}(\gamma) \equiv {\bf u}^{(k)}(\gamma+\frac{\pi}{2})$.
Once the basis, i.e. $\gamma$, is chosen, the matrix $\bS^{(\gamma)}$ which
diagonalizes $\A$ is given by
\beq
S_{nk}^{(\gamma)}=  w^{(k)}_{n}(\gamma)~,~~~~~~
\begin{array}{l} n=1,\ldots,N \\
 k=0,\ldots,N-1~. \end{array}
\label{S}
\eeq
Using this matrix in the generating function of the form~(\ref{generating}),
the set of normal coordinates and momenta
$(\Q^{(\gamma)},\bP^{(\gamma)})$, associated to the chosen basis,
is defined by the canonical linear orthogonal transformation
\beq
\left\{ \begin{array}{lll}
\q & = &\bS^{(\gamma)}\Q^{(\gamma)} \\
\p & = &\bS^{(\gamma)}\bP^{(\gamma)}~.
\end{array} \right.
\label{QPmapstoqp}
\eeq
For each fixed $k$, the set of components
$\{w^{(k)}_{n}(\gamma)\}_{n\in\{1,\ldots,N\}}$, viewed as a function of
the lattice position $n$, gives the spatial pattern associated
to the presence of the normal coordinate $Q_{k}^{(\gamma)}$, since
$q_{n}(t) = \sum_{k=0}^{N-1} Q_{k}^{(\gamma)}(t)  w^{(k)}_{n}(\gamma)$.
Choosing a basis amounts to a choice of the spatial phase factor $\gamma$
for the stationary basis waves used to Fourier analyze the spatial
pattern of the chain, given by the set $\{q_{n}\}$.
Different basis give different Fourier components, i.e. different
sets of normal coordinates.
The coordinate $Q_{0}$ is proportional to the displacement of the system's
center of mass, while for $k \in \{ 1,\ldots,[\frac{N-1}{2}] \}$
both coordinates $(Q_{k}^{(\gamma)},Q_{N-k}^{(\gamma)})$
correspond, in the Fourier analysis of the spatial pattern,
to the presence of a component of wavelength $N/k$
(the lattice spacing being the unit of length)
but they refer respectively to a sine and cosine wave
(with respect to the chosen spatial phase).
Finally the $Q_{N/2}$ coordinate corresponds to the pattern of wavelength two,
where adjacent particles have equal but opposite displacements.

Every choice of the basis is equivalent
{\em as far as the linear system is concerned\/}
since Hamiltonian~(\ref{H0}) takes the same form in every set of
normal coordinates $(\Q^{(\gamma)},\bP^{(\gamma)})$.
This is no longer true for the full nonlinear Hamiltonian $H$,
when the same canonical transformation is applied.

\subsection{Transformation of the full Hamiltonian}
\label{Transformation}

We now choose once for all the basis with $\gamma=\pi/4$.
This is done for two reasons. Firstly, this choice simplifies the
calculations since all the elements of the matrix $\bS \equiv \bS^{(\pi/4)}$
are given by the single expression
\beq
S_{nk} \equiv w^{(k)}_{n}(\pi/4) =
\frac{1}{\sqrt{N}}
\left[ \sin \left( \frac{2\pi kn}{N} \right)+
\cos \left( \frac{2\pi kn}{N} \right) \right]
\, ,~~
\begin{array}{l} n=1,\ldots,N    \\
k=0,\ldots,N-1  \end{array}
\label{defS}
\eeq
Moreover, we shall see that some of the normal modes of this basis
can be ``extended'' to the anharmonic case.
 From now on we denote simply $(\Q,\bP) \equiv (\Q^{(\pi/4)},\bP^{(\pi/4)})$.

The quartic term $H_1$ in~(\ref{H1(q)}) is transformed into
\beq
H_{1}(\Q,\bP) = \frac{\beta}{4}\sum_{n=1}^{N} \left( \sum_{k=0}^{N-1}
(S_{n+1,k}-S_{nk}) Q_{k} \right)^{4} ~,
\label{H1(Q)}
\eeq
with $S_{N+1,k} \equiv S_{1,k}$.

Since
\beq
S_{n+1,k}-S_{nk}=\sqrt{\frac{2}{N}} \, \omega_{k}
\cos \left( \frac{\pi k(2n+1)}{N}+\frac{\pi}{4} \right)~,
\eeq
it is useful to introduce the quantities
\beq
U_{nk}^{\delta} \equiv
\cos \left( \frac{\pi k(2n+1)}{N}+\delta \right) ~,
\eeq
which obey the following rules
\bea
U_{nr}^{\phi}U_{ns}^{\delta} & = & \frac{1}{2}
\left( U_{n,r+s}^{\phi+\delta}+U_{n,r-s}^{\phi-\delta} \right)~,
\label{algebra}\\
U_{nr}^{\delta+\pi} & = & -U_{nr}^{\delta}~.
\label{piupi}
\eea
The nonlinear term~(\ref{H1(Q)}) can be rewritten in terms of
these quantities as
\beq
H_{1}(\Q)=\frac{\beta}{N^{2}} \sum_{i,j,k,l=1}^{N-1}
\omega_{i}\omega_{j}\omega_{k}\omega_{l}Q_{i}Q_{j}Q_{k}Q_{l}
\sum_{n=1}^{N} U_{ni}^{\pi /4}U_{nj}^{\pi /4}U_{nk}^{\pi /4}U_{nl}^{\pi /4}~.
\label{H1-2}
\eeq
Using the properties~(\ref{algebra}),(\ref{piupi}) we get
\bea
\lefteqn{\sum_{n=1}^{N}
U_{ni}^{\pi /4}U_{nj}^{\pi /4}U_{nk}^{\pi /4}U_{nl}^{\pi /4} = }  \nonumber\\
&&\frac{1}{8}\sum_{n=1}^{N} \left(
-U_{n,i+j+k+l}^{0}+U_{n,i+j-k-l}^{0}+U_{n,i-j+k-l}^{0}+U_{n,i-j-k+l}^{0}
\right)+ \nonumber \\
&&+\frac{1}{8}\sum_{n=1}^{N} \left(
+U_{n,i+j+k-l}^{\pi/2}+U_{n,i+j-k+l}^{\pi/2}+U_{n,i-j+k+l}^{\pi/2}
-U_{n,i-j-k-l}^{\pi/2} \right)~.
\label{prodU}
\eea
Using the identity, valid for integer $r$,
\beq
\sum_{n=1}^{N} \exp\left( i \frac{\pi r(2n+1)}{N} \right) =
\left\{ \begin{array}{ll}
(-1)^{m} N & \mbox{if $r=mN$ with } m \in \mbox{\boldmath $Z$} \\
0 & \mbox{otherwise~,}
\end{array} \right.
\eeq
we obtain
\beq
\sum_{n=1}^{N} U_{n,r}^{\pi/2}=0~~~,\forall r \in \mbox{\boldmath $Z$}~,
\label{Upi/2}
\eeq
and
\beq
\sum_{n=1}^{N} U_{n,r}^{0}=N \Delta_{r}~~~,
\forall r \in \mbox{\boldmath $Z$}~, \label{U0}
\eeq
being
\beq
\Delta_{r}=\left\{ \begin{array}{ll}
(-1)^{m} & \mbox{for $r=mN$ with } m \in \mbox{\boldmath $Z$} \\
0 & \mbox{otherwise~.}
\end{array} \right.
\label{delta}
\eeq
 From~(\ref{H1-2}), (\ref{prodU}), (\ref{Upi/2}) and~(\ref{U0}) we get
\beq
H_{1}(\Q)= \frac{\beta}{8N} \sum_{i,j,k,l=1}^{N-1}
\omega_{i}\omega_{j}\omega_{k}\omega_{l} C_{ijkl}
 Q_{i}Q_{j}Q_{k}Q_{l}~,
\label{H1fin}
\eeq
with
\beq
C_{ijkl}=-\Delta_{i+j+k+l}+\Delta_{i+j-k-l}+\Delta_{i-j+k-l}+\Delta_{i-j-k+l}
\label{Cijkl}
\eeq
These integer coefficients are invariant under any permutation of the indices.

The full Hamiltonian of the FPU $\beta$-model for $N$ oscillators
with periodic boundary conditions in the new coordinates is
\beq
H(\Q,\bP) = \frac{1}{2} P_{0}^2 +
\frac{1}{2} \sum_{i=1}^{N-1} (P_{i}^2+\omega_{i}^{2}Q_{i}^2) + H_{1}(\Q)~.
\label{Hfin}
\eeq
Eliminating center of mass motion, we get easily also the equations of
motion for the remaining $N-1$ degrees of freedom, which, in second order
form, read
\beq
\left\{ \begin{array}{lll}
\ddot{Q}_r & = & F_r (Q_1,\ldots,Q_{N-1})~,~~~~~~~r=1,\ldots,N-1 \\
{F}_r (Q_1,\ldots,Q_{N-1})  & = & {\displaystyle -\omega_{r}^{2}Q_{r} -
\frac{\beta\omega_{r}}{2N} \sum_{j,k,l=1}^{N-1}
\omega_{j}\omega_{k}\omega_{l}
C_{rjkl} Q_{j}Q_{k}Q_{l} }~,
\end{array} \right.
\label{eqmoto}
\eeq
where $F_r$ is the generalized force in normal coordinate space.

\section{Intermezzo: the integrable $N=3$ case}
\label{Integrable}

It is well known that the FPU $\beta$-model with $N=3$ and periodic boundary
conditions is integrable~\cite{Chood}. This is due to the presence
of a third independent integral
of motion besides energy and total momentum.
In ref.~\cite{Chood} it is in fact shown that the quantity
\beq
M(\q,\p) =
p_{1} (q_{2}-q_{3}) + p_{2} (q_{3}-q_{1}) + p_{3} (q_{1}-q_{2})
\label{M}
\eeq
is a constant of the motion.
Our canonical form of the FPU Hamiltonian~\mbox{(\ref{H1fin},\ref{Hfin})}
allows a simple interpretation
of this new integral and clarifies why it is typical of $N=3$.

In the case $N=3$ there are only two non-zero harmonic frequencies,
the coincident pair $\omega_1=\omega_2= \sqrt{3}$.
Invariance under permutation of indices of the
$C_{ijkl}$ coefficients in formula~(\ref{H1fin})
allows to restrict the search for non zero
coefficients among those with
$i \geq  j \geq  k \geq l$. These are $C_{1,1,1,1}=C_{2,2,2,2}=3$
and $C_{2,2,1,1}=1$.
The resulting Hamiltonian is
\beq
H(\Q,\bP)=\frac{1}{2} P_{0}^2 +
\frac{1}{2} \left( P_{1}^2+P_{2}^2 \right)+
\frac{3}{2} \left( Q_{1}^2+Q_{2}^2 \right)+
\frac{9}{8} \beta \left( Q_{1}^2+Q_{2}^2 \right)^{2}~.
\label{N=3}
\eeq
Obviously the center of mass motion is free, and $P_0$ is conserved.
More important is the invariance of the potential term in~(\ref{N=3})
under rotations in the $(Q_1,Q_2)$ plane,
which is transparent in these new coordinates.
This means
that the {\it pseudo-angular momentum} $L=(Q_1P_2-Q_2P_1)$ is conserved.
Going back to old coordinates and momenta by the inverse of the
transformation~(\ref{QPmapstoqp}), we get easily the new integral in~(\ref{M}):
\beq
M = \sqrt{3}\, L~.
\eeq
Thus, the presence of this new integral is not a surprise since it
arises from a simple symmetry which was not apparent in the original
coordinates; this symmetry is no more present for $N >  3$,
since the potential does not have in general rotational invariance in the
planes $(Q_k,Q_{N-k})$, but it holds true in the invariant subspace
$\{N/3,2N/3\}$ (see Subsection~\ref{Two-mode}).

We finally observe that one can give a different interpretation
of the system considered in this section.
Let us think of the coordinates $(q_{1},q_{2},q_{3})$ as cartesian
coordinates of a single particle governed by the Hamiltonian~(\ref{H})
with $N=3$,
thus identifying the configuration space of the three one-dimensional
oscillators system with that of a single particle in three-dimensional space.
The transformation to normal coordinates, being induced by an orthogonal
matrix, is a rotation of the reference frame in this context,
and the form~(\ref{N=3})
taken by the hamiltonian in these new coordinates reveals the cylindrical
symmetry of the potential around the ${\bf w}^{(0)}$ axis and its
independence from the $Q_{0}$ coordinate associated to the latter.
 From this immediately follows the conservation law for $L$, the angular
momentum component along the ${\bf w}^{(0)}$ axis,
and for $P_0 = (p_{1}+p_{2}+p_{3})/\sqrt{3}$ which is now interpreted
as the particle's momentum component $\p \cdot {\bf w}^{(0)}$;
in fact there is no force along that direction.

We also observe that nothing changes if the quadratic part of the potential
in~(\ref{N=3}) is omitted, so that also the purely quartic periodic chain
with three particles is an integrable system.

\section{Explicit low-dimensional solutions}
\label{Explicit-solutions}

 From the expression~(\ref{Cijkl}) of the numerical coefficients
appearing in the
generalized force $F_r$ in~(\ref{eqmoto}) it is easy to derive some
selection rules
already empirically found in the literature.
For instance, it is known (see e.g.~\cite{Kantz89,Livi83-exponential}) that
for a periodic FPU-$\beta$ chain with an even number of oscillators
the initial excitation of
a set of modes all having even (resp. odd) indices only, cannot lead
to the excitation of modes having odd (resp. even) indices.
This property can be easily derived
observing that initial conditions of these two kinds assign a special role,
among all the coefficients $C_{rjkl}$, to those with one index of a
given parity  and all the others of the opposite parity.
For example, if even modes only are initially excited,
it follows from~(\ref{eqmoto})
that odd modes can be excited only through those $C_{rjkl}$
with only one odd index;
the converse applies to the initial excitation of odd modes only.
Since the algebraic sums of indices of the $\Delta$-symbols defining
the coefficients in~(\ref{Cijkl}) are,
in both cases, always odd integers, they cannot be multiples of
the even number $N$.
Thus, all $C_{rjkl}$ connecting (in the above sense)
odd to even modes vanish.

This example shows that (for even $N$), if we consider the set
of all modes partitioned in the two subsets of the even and odd modes,
this partition has the property that an initial excitation
completely contained in one of the two subsets cannot propagate
to the other. It is therefore natural to wonder about the existence of
other partitions of the set of modes having the same property.
Exploiting this idea we are able to find some explicit
periodic and quasiperiodic solutions for the FPU-$\beta$ system.
Though their existence, for a finite number of oscillators $N$,
is strongly dependent on the divisibility properties of $N$,
once the solutions are obtained and expressed in terms of the original
coordinates $\{q_{n}\}$ they remain valid for the infinite chain.

Let us denote with $\cal M$ the set of all ``internal''
modes for an arbitrarily fixed number of particles $N$,
i.e. the set of indices  \mbox{${\cal M} \equiv \{1,\ldots,N-1\}$}.
The previous example leads us to introduce the following definition:
a subset of modes ${\cal A} \subset {\cal M}$ is defined
to be ``of type I'' when
\beq
C_{ijkl}
=0~~~~~\forall i \in {\cal M}\setminus{\cal A}\,,~~\forall j,k,l \in {\cal A}~.
\label{type-I}
\eeq
It is clear that, if a subset ${\cal A}$ is of type I, every solution
of~(\ref{eqmoto}) having
$Q_i(0)=\dot{Q_i}(0)=0$, $\forall i \in {\cal M}\setminus{\cal A}$, is such
that
$Q_i(t) \equiv 0$ $ \forall t$ for the same indices $i$, which means that
an initial excitation
completely contained in ${\cal A}$ cannot propagate out of ${\cal A}$.
The remaining coordinates $\{Q_j(t)\}_{j \in \cal A}$ of the solution
obey a reduced system of equations of motion analogous to~(\ref{eqmoto}),
with the only difference that the summation in the expression of
the generalized force is restricted
to indices $j,k,l \in \cal A$.
So the normal coordinates and conjugated momenta with indices
in ${\cal A}$
span an invariant subspace in the system's phase space,
with dimension given by the double of
the elements of ${\cal A}$; the dynamics over it is generated
by the reduced Hamiltonian
\bea
H_{\cal A}(\{Q_j,P_j\}_{j \in \cal A}) & = &
\frac{1}{2} \sum_{i \in \cal A} (P_{i}^2+\omega_{i}^{2}Q_{i}^2) + \nonumber \\
 & & \frac{\beta}{8N} \sum_{i,j,k,l \in \cal A}
\omega_{i}\omega_{j}\omega_{k}\omega_{l}
C_{ijkl} Q_{i}Q_{j}Q_{k}Q_{l}~.
\label{Hrid}
\eea

The subsets of even and odd modes (for even $N$) are just two examples
of subsets of type I (each one with the additional non-generical peculiarity
of having the complementary subset also of type I). It is easy to prove
that these two subsets are type I independently of the chosen basis~(\ref{w}).

In the following subsections we discuss some low-dimensional type I
subsets that we were able to identify, and their associated exact solutions.
It must be remembered that we refer to the normal coordinates with
$\gamma = \pi/4$ in~(\ref{QPmapstoqp}).

\subsection{One-mode solutions}
\label{One-mode}

Let us begin with the case of the simplest subsets ${\cal A}$:
those consisting of only one mode, denoted by the index $e$.
Then, the condition~(\ref{type-I}) for ${\cal A} \equiv \{e\}$
to be of type I is
\beq
C_{neee} = 0~~~~~\forall n \in {\cal M} \mbox{ with } n \neq e~.
\label{one-mode-condition}
\eeq
If this happens for some $e$, the corresponding reduced
Hamiltonian~(\ref{Hrid}) represents a one degree of freedom
(and thus integrable) system, described by the single coordinate $Q_e$.

Writing the required coefficients in terms of the $\Delta$-symbols
of~(\ref{delta})
\beq
C_{neee} = - \Delta_{n+3e} + 3 \Delta_{n-e}~,
\label{cne}
\eeq
we find that for each fixed $e \in \cal M$, except for
\beq
e=\frac{N}{4};\: \frac{N}{3};\: \frac{N}{2};
\: \frac{2N}{3};\: \frac{3N}{4}~,
\label{solitari}
\eeq
there is in $\cal M$ always one (and only one) $n \neq e$ such that
$C_{neee} \neq 0$; thus condition~(\ref{one-mode-condition})
is not satisfied and $\{e\}$ is not of type I.
Given a mode $e \in \cal M$ different from those listed in~(\ref{solitari}),
the unique $n \in \cal M$ different from $e$ such that $C_{neee} \neq 0$
is in fact given by the
unique integer in $\cal M$ congruent to $-3e$ modulo $N$.
Such a mode
can be called the mode $n=\bar{n}(e)$ ``directly forced'' by mode $e$, since
in the generalized force $F_{\bar{n}}$ in~(\ref{eqmoto}) a forcing term
proportional to $C_{\bar{n}eee} Q_{e}^{3}$ is present.
This does not mean that this mode
will be the fastest growing mode, when mode $e$ is the only one
initially excited
(since, of course, resonance relations must be examined), but its role
is important because it acts as a trigger to excite other modes $Q_j$
(possibly more unstable but not directly forced by mode $e$)
through terms of the kind $C_{j\bar{n}ee}Q_{\bar{n}}Q_{e}^{2}$
in their equations of motion~(\ref{eqmoto}).

On the contrary, for each of the modes listed in~(\ref{solitari})
(of course when some of them exists, i.e. when $N$ has the divisibility
property required for the considered $e$ in~(\ref{solitari}) to be an integer)
property~(\ref{one-mode-condition}) is verified; thus these are the
only values of $e$ such that $\{e\}$ is a type I one-mode subset.
If one of the modes in~(\ref{solitari}) is the only mode initially excited,
it remains excited without transferring energy to any other mode, and the
only modes that possess this property are those in~(\ref{solitari}).
In this case, the equation of motion for the excited mode amplitude $Q_e$ is
\beq
\ddot{Q}_{e} = - \omega_e^2 Q_{e} -
\frac{\beta \omega_{e}^4 C_{eeee}}{2 N} Q_{e}^3~,
\label{eqe}
\eeq
where $C_{eeee}$ is always positive.
The general solution of~(\ref{eqe}), having as free parameters
the amplitude $A$ and the time origin $t_0$, is
\beq
Q_{e} (t) = A \: \cn \left[ \Omega_e (t -t_0) \, ,\, k  \right]~,
\label{e}
\eeq
where $\Omega_e$ and the modulus $k$ of Jacobi elliptic cosine
function\footnote{For standard notation and properties of elliptic functions
and integrals we refer to~\cite{Lawden,Abramowitz}.}
both depend on $A$:
\bea
\Omega_e & = & \omega_e  \sqrt{1+\delta_e  A^2} \label{Omega} \\
k & = & \sqrt{\frac{\delta_e A^2}{2 (1 +\delta_e A^2)}}~~, \label{modulus}
\eea
with $\delta_e=\beta \omega_e^2 C_{eeee}/(2N)$.
We remind that this solution is periodic, with amplitude
dependent period $T_e$ given in terms of
the complete elliptic integral of the first kind $K(k)$ by
\beq
T_e = \frac{4 K(k)}{\Omega_e}~.
\label{period}
\eeq
Since $A$ is one-to-one related to the control parameter~(\ref{epsilon})
involving the energy density, all parameters of the solution~(\ref{e})
can be rewritten as functions of $\epsilon$ instead of $A$.
For the period we have
\beq
T_e = \frac{4 K(k)}{\omega_e (1 + 2 \epsilon C_{eeee})^{1/4}}~,
\label{period2}
\eeq
where $k=k(\epsilon)$.
The period decreases as
$\epsilon$ is increased, while in the limit
$\epsilon \rightarrow 0$ it reduces to the
harmonic period $T_e \rightarrow 2 \pi / \omega_e$.

All one-mode solutions correspond to stationary waves of the form
\beq
q_n (t) = Q_e (t) w_n^{(e)}
\eeq
where the vectors ${\bf w}^{(e)} \equiv {\bf w}^{(e)} (\pi/4)$
are given in formula~(\ref{defS}). The spatial pattern is therefore
the same of the corresponding linear mode, but these are {\em nonlinear
modes\/}
since the time dependence corresponds to a nonlinear oscillation,
which reduces to the usual harmonic one as $A$ goes to zero.

Let us first discuss the lattice pattern of  the solution corresponding
to $e=N/2$, i.e. its expression
as a function of the old set of coordinates:
\beq
q_n(t) = \frac{(-1)^n}{\sqrt{N}} Q_{N/2} (t)~,
\label{N/2old}
\eeq
where $Q_{N/2}(t)$ is given by Eqs.~(\ref{e}), (\ref{Omega})
and~(\ref{modulus}),
with $e=N/2$, $\omega_{N/2}=2$, $\delta_{N/2}=4\beta/N$.
The lattice pattern has adjacent particles with opposite phases,
being the same of the zone-boundary phonon mode.
Actually, this solution is present for any potential of the form
\beq
V= \sum_{n=1}^N \Phi (q_{n+1}-q_n)~.
\eeq
with even $N$ on a periodic lattice (or on an infinite lattice).
In fact, after the substitution of the ansatz solution $q_n = (-1)^n r(t)$,
the equations of motion reduce to the single
equation for $r(t)$
\beq
\ddot{r} =  - 2 \frac{d \Phi_{even}}{dr}(2r)~,
\eeq
where $\Phi_{even}(r) \equiv (\Phi(r)+\Phi(-r))/2$
is the even part of the $\Phi$ function.
For instance,
a solution of this kind is known for the integrable case of the
Toda lattice and corresponds to the standing cnoidal wave with
two lattice spacings wavelength~\cite{Toda}.

The cases $e=N/4$ and $e=3N/4$  are treated in a similar way.
The lattice pattern is
\beq
q_n(t) = \frac{1}{\sqrt{N}} \, Q_e (t) \left[
\pm \sin \left(\frac{\pi n}{2}\right) +
\cos \left(\frac{\pi n}{2}\right) \right]
\label{N/4old}
\eeq
where $Q_e (t)$ is given by Eqs.~(\ref{e}), (\ref{Omega}) and~(\ref{modulus})
with $\omega_{N/4}=\omega_{3N/4}=\sqrt{2}$,
$\delta_{N/4}=\delta_{3N/4}=4 \beta/N$, and the `$\,+\,$' and `$\,-\,$' signs
correspond to $e=N/4$ and $e=3N/4$ respectively.
Let us observe that the solution $e=3N/4$ is obtained from that with $e=N/4$
by shifting the lattice pattern by
one lattice spacing; another shift produces the same solution apart from a
temporal phase.
This property is a consequence of the invariance of the model under
translations of an integer number of lattice spacings.

Because of an additional peculiarity, the single mode solutions $e=N/3$
and $e=2N/3$
will be treated in the following as particular cases of the two-mode manifolds.

\subsection{Two-mode manifolds}
\label{Two-mode}

We do not try here to identify all multi-mode, or even two-mode, type I subsets
for generic $N$, although some of them can be found by simple inspection
of a computed table of non-zero $C_{ijkl}$
coefficients\footnote{For example, it can be seen that $\{1,5\}$ and $\{3,7\}$
for $N=8$, or $\{1,5,9\}, \{2,6,10\}$ and $\{3,7,11\}$ for $N=12$
are all of type I.} for small $N$.
Instead we confine ourselves to the search of two-mode type I subsets
of the kind $\{i,N-i\}$, since in this case we get complete results
leading to periodic and quasiperiodic explicit solutions.

A subset $\{i,N-i\}$ is of type I when (see condition~(\ref{type-I}))
\beq
C_{njkl}
=0~~~~~\forall n \in {\cal M}\setminus\{i,N-i\},~\forall j,k,l \in \{i,N-i\}~.
\label{two-mode-condition}
\eeq
Thus, in particular, a {\em necessary} condition is
\beq
C_{niii} = 0~~~~~\forall n \in {\cal M} \mbox{ with } n \neq i,N-i~.
\eeq
This can happen, {\em a priori}, only in two cases:
{\it a)} $\{i\}$ is of type I,
or {\it b)} $\{i\}$ is not type I but the mode $\bar{n}(i)$,
directly excited by $i$, is coincident with $N-i$. But it is easy to see,
 from the explicit expression of $\bar{n}(i)$,
that case {\it b)} can never happen: then if $i$ is such that
$\{i,N-i\}$ is of type I, it must be one of the modes in~(\ref{solitari}).
The only subsets of the kind requested that can be type I are therefore
${\cal E}_1 \equiv \{ N/4,3N/4 \}$ and ${\cal E}_2 \equiv \{ N/3,2N/3 \}$.
To check that they are indeed, it is useful to remark that
\beq
C_{N-i,\,N-j,\,N-k,\,N-l}=C_{ijkl}~.
\label{C(N-i)}
\eeq
Then, since $i$ verifies~(\ref{one-mode-condition}) with $e=i$
(and also $N-i$ does), from~(\ref{C(N-i)}) we need only to check that
\beq
C_{n,\,i,\,i,\,N-i}=0~~\forall n \in {\cal M}\setminus\{i,N-i\}
\label{check}
\eeq
for condition~(\ref{two-mode-condition}) to be satisfied.

Let us first consider the set ${\cal E}_1$ .
Using formula~(\ref{Cijkl}) we have
\beq
C_{n,\, N/4,\, N/4,\, 3N/4}=-\Delta_{n+5N/4}+2\Delta_{n-3N/4}+\Delta_{n+N/4}~,
\label{CE1}
\eeq
and each $\Delta$-symbol vanishes for $n \not\in  {\cal E }_1$, thus
satisfying~(\ref{check}).
This proves that
the set ${\cal E}_1$ is of type I, i.e. the excitation of the the pair
$\{ N/4, 3N/4 \}$ does not propagate to other modes.
Moreover,
Hamiltonian~(\ref{Hrid}) with ${\cal A}={\cal E}_1$, describing the reduced
system containing only this pair of modes, is separable because
the coefficients~(\ref{CE1}) are zero also for $n \in {\cal E}_1$, giving
\beq
H_{{\cal E}_{1}} = \frac{1}{2} \left(  P_{N/4}^2+P_{3N/4}^2 \right)+
Q_{N/4}^2+Q_{3N/4}^2+
\frac{2\beta}{N} \left( Q_{N/4}^4+Q_{3N/4}^4 \right)
\label{E1}
\eeq
Being separable, this Hamiltonian is integrable. Thus,
the motion in the subspace of the modes $\{N/4,3N/4\}$
is simply given by the cartesian product of the one-mode solutions already
introduced in the previous subsection.
However, it is interesting to observe that this two-mode
invariant manifold foliated in two-dimensional tori exists
in the $2N$-dimensional phase space of the system,
with periodic or quasiperiodic trajectories whose winding number depends on
the non-linear periods of the two modes (see formula~(\ref{period})).
Therefore we are able to obtain
quasiperiodic solutions with two frequencies exploiting the fact that
anharmonicity, causing
the nonlinear dependence on amplitude of the period~(\ref{period}),
``removes'' the harmonic frequency degeneracy of the pair $\{N/4,3N/4\}$.
The general solution (see formula~(\ref{e})) is
\bea
\lefteqn{ q_n (t) = }  \label{N43N4}  \\
 & & \frac{1}{\sqrt{N}} A_{N/4} \,
\cn \left[ \Omega _{N/4}(A_{N/4}) \,t + \phi,\, k(A_{N/4}) \right]
\left[ \sin \left( \frac {\pi}{2} n \right) +
\cos \left(\frac {\pi}{2} n \right) \right] + \nonumber \\
 & & \frac{1}{\sqrt{N}} A_{3N/4} \,
 \cn \left[ \Omega_{3N/4} (A_{3N/4}) \,t + \psi, \, k(A_{3N/4}) \right]
\left[ - \sin \left(\frac {\pi}{2} n \right) +
\cos \left(\frac {\pi}{2} n \right)  \right]~, \nonumber
\eea
where we have emphasized the dependence of the various parameters
on the two independent arbitrary amplitudes $A_{N/4}$ and $A_{3N/4}$.
It is interesting to observe that
\beq
q_{n+2} = - q_n~,
\eeq
which explains why a mode with a wavelength of  four lattice sites
corresponds to a solution with two degrees of freedom.
When the amplitudes of the two modes in~(\ref{N43N4}) are equal
(we omit in this case the moduli in
the arguments of the cn functions since they are equal)
one gets periodic solutions. In particular, for
the two choices of the phase difference $\psi - \phi = 0$ and
$\psi - \phi = 2K$, stationary wave solutions are obtained. They correspond
to a lattice pattern having one node every two lattice sites,
thus giving also a solution valid for fixed end chains,
after an appropriate choice of the lattice origin. It should also be
observed that these latter would be one-mode solutions in the normal
coordinates relative to the basis~(\ref{w}) with $\gamma=0$.

If instead we choose $\psi - \phi = \pm K$ [still with equal amplitudes,
denoted by $A$, and with $K \equiv K(k(A))$] the corresponding periodic
solutions have the property
\beq
q_{n+1} (t \pm \Delta t) = q_n (t)~,~~~
\mbox{for $\Delta t =  K/ \Omega_{N/4}(A)~,\forall n$}~,
\label{twp}
\eeq
where the sign in the argument in the l.h.s. is the same of $(\psi - \phi)$.
We consider~(\ref{twp}) as the defining property of a travelling wave
on a lattice, propagating respectively to the right (`$\,+\,$' sign)
or to the left (`$\,-\,$' sign) with velocity $1/ \Delta t$
(the lattice spacing being the unit of length). We remark that the
velocity depends on the amplitude, as should be expected for a
nonlinear wave. Actually, observing that, for integer $n$,
the following identities hold
\bea
\sin \left( \frac {\pi}{2} n \right) \cn (x) &=&
\frac{1}{2} \left\{ \cn [ K (n-1) + x ] + \cn [K(n-1) -x] \right\}~,
\label{ident1}\\
\cos \left( \frac {\pi}{2} n \right) \cn (x) & = &
\frac{1}{2} \left\{ \cn ( K n + x ) + \cn (Kn -x) \right\}~,
\label{ident2}
\eea
and defining the function
\[
\mbox{g} (x,k) \equiv \cn (x-K(k),\,k) + \cn (x,k)~,
\]
it is straightforward to rewrite the general solution~(\ref{N43N4}),
for excitations in the ${\cal E}_1$ subspace,
in a form which
reveals its progressive and regressive travelling wave content,
in analogy with the harmonic case:
\bea
\lefteqn{q_n (t) =}       \label{sovrapposiz}  \\
 & & \frac{1}{2 \sqrt{N}}
\left[A_a \mbox{g} (K_a n + \Omega_a t + \phi, \,k_a) +
A_{b} \mbox{g} (K_{b}(n+1) + \Omega_{b} t + \psi,\,k_{b}) \right] +
\nonumber \\
 & & \frac{1}{2 \sqrt{N}}
\left[A_a \mbox{g} (K_a n - \Omega_a t - \phi, \,k_a) +
A_{b} \mbox{g} (K_{b}(n+1) - \Omega_{b} t - \psi,\,k_{b}) \right]~.
\nonumber
\eea
In~(\ref{sovrapposiz}) we have denoted the quantities relative to the
modes $N/4$ and $3N/4$ by the indices `$a$' and `$b$' respectively.
The particular cases above, with $A_a=A_{b}=A$
and $\psi - \phi = \pm K$ are
\beq
q_n (t) = \frac{A}{\sqrt{N}} \,
\mbox{g}  \left( K n \mp \Omega_{N/4} (A) \, t \mp \phi \right)~,
\label{g-travelling}
\eeq
where we have made explicit the travelling wave nature of these solutions.

Turning now to the other set ${\cal E}_2$, we have
\beq
C_{n,\, N/3,\, N/3,\, 2N/3} =
-\Delta_{n+4N/3} + 2 \Delta_{n-2N/3} + \Delta_{n}~,
\eeq
and again we find that all $\Delta$'s vanish for $n \not\in {\cal E}_2$,
thus proving that also ${\cal E}_2$ is of type I.
This time there is a non-zero case among the different kinds
of coefficients coupling the two modes of this
subset, e.g. $C_{N/3,\, N/3,\, 2N/3,\, 2N/3}=1$, and the reduced
Hamiltonian is
\beq
H_{{\cal E}_{2}} = \frac{1}{2} \left(  P_{N/3}^2 + P_{2N/3}^2 \right)+
\frac{3}{2} \left(  Q_{N /3}^2 + Q_{2N/3}^2 \right) +
\frac{27\beta}{8N} \left( Q_{N/3}^2 + Q_{2N/3}^2 \right)^2~.
\label{E_2}
\eeq

Let us first remark that this Hamiltonian reduces to the one
of formula~(\ref{N=3}) for $N=3$
apart from the center of mass motion term, which has already been eliminated.
Being the potential in~(\ref{E_2}) invariant under rotation in the
$ (Q_{N/3}, Q_{2N/3})$ plane, again, as for the $N=3$ case,
the ``angular momentum'' $(Q_{N/3}P_{2N/3} - Q_{2N/3}P_{N/3})$ is conserved.
Thus,  having two constants of motion in involution,
Hamiltonian~(\ref{E_2}) is integrable.
We have again an invariant manifold foliated by two-dimensional tori,
but the explicit general solution for Hamiltonian~(\ref{E_2}),
although possible in principle, is less trivial in this case.
However, we can easily find some lower dimensional solutions.
In fact, the presence of a central potential implies that there exist
solutions whose trajectory on the $(Q_{N/3},Q_{2N/3})$ plane is an
oscillation lying on an arbitrary chosen straight line crossing the origin.
Hence,  this implies that not only the modes $N/3$ and $2N/3$ are
particular one-mode solutions,
as we have already anticipated, but that there are analogous solutions
corresponding to all possible choices (rotations) of the basis~(\ref{w});
this property is not present for the set ${\cal E}_1$.
Each one of these solutions with zero ``angular momentum''
would be regarded as one-mode
if we had chosen a linear basis with the convenient value of $\gamma$
in formula~(\ref{w}). In the original lattice coordinates $\{ q_n \}$
they are stationary waves, with a pattern given by an arbitrary continuous
translation of the stationary sine wave of wavelength three.
All these one-mode solutions satisfy the same differential
equation~(\ref{eqe}), with $\omega_{e}=\omega_{N/3}=\sqrt{3}$ and
$C_{eeee}=C_{N/3,\, N/3,\, N/3,\, N/3}=3$.
The lattice pattern is
\beq
q_n (t) = \sqrt{\frac{2}{N}} \, A \,
\cn \left[\Omega_{N/3} (t-t_0), \, k \right]
\sin \left(\frac{2\pi}{3} n + \gamma \right)~,
\label{N3gamma}
\eeq
where $\gamma$ is an arbitrary real parameter;
$\Omega_{N/3}$ and $k$ are given by~(\ref{Omega}) and~(\ref{modulus})
with $\delta_e=\delta_{N/3}=9 \beta/(2N)$.
It is interesting to observe that for a generic value of $\gamma$
this solution has no nodes
(still oscillators) at lattice sites. The only case in which nodes appear
is $\gamma = 0$
(the cases $\gamma = \pi /3$ and $\gamma = 2 \pi/3$ can be reduced
to this one after
a translation of a lattice unit), and it is clearly present
also for fixed ends oscillator chains.

Also for the set ${\cal E}_2$ travelling wave solutions exist.
They correspond to closed orbits for
the central potential in Hamiltonian~(\ref{E_2}).
A particularly simple case is the one of closed
circular orbits:
\beq
\left\{
\begin{array}{lll}
Q_{N/3}(t)& =& A \cos  \left[ \Omega_{N/3} (t - t_0) \right] \\
Q_{2N/3}& =& \pm A \sin  \left[ \Omega_{N/3} (t - t_0) \right]~.
\end{array}
\right.
\eeq
The lattice pattern is
\beq
q_n (t) = \sqrt{\frac{2}{N}} \, A \,
\sin \left( \frac{2 \pi}{3} n \pm \Omega_{N/3} \, t +\phi \right)
\label{tw2}
\eeq
where $\phi$ is an arbitrary phase.
This solution is of the form of the linear travelling wave, but
has amplitude dependent frequency and velocity,
since $\Omega_{N/3}=\Omega_{N/3}(A)$.

Finally, we further stress that, though for the FPU-$\beta$ system
with finite $N$
each of these solutions exists only for suitable (but still infinitely many)
values of $N$, their expressions in terms of the original $\{q_n\}$ coordinates
remain valid as spatially periodic solutions for the infinite chain.
The ``out-of-phase'' oscillation~(\ref{N/2old}) and the one-mode
$N/4$ solution with $\gamma=0$ (given by our solution~(\ref{N43N4}) in the
particular case $A_{N/4}=A_{3N/4}$, $\psi=\phi$) appeared first
in~\cite{Budinsky}, where they were obtained directly in particle
displacement coordinates by a suitable ``ansatz''.
A statement equivalent, in our notation, to the fact
that an excitation in the subsets ${\cal E}_1$ or ${\cal E}_2$ cannot
propagate out of these sets, was given in unpublished work
by Sholl~\cite{Sholl-tesi}.

\section{Stability of periodic orbits}
\label{Stability}

In the previous Section we have shown the existence,
in the phase space of the FPU-$\beta$ system, of various kinds
of initial conditions giving rise to trajectories lying on manifolds
of smaller dimension than that of the constant energy shell.
We have considered in detail some simple cases leading to reduced
integrable systems, thus obtaining some families of explicit solutions.
In our opinion these are interesting in their own, since explicit analytical
solutions, are not common at all for nonlinear {\em non-integrable} chains
of oscillators. However we must remember that all these results hold
for {\em particular} classes of initial conditions. Therefore an important
problem is that of the stability of these solutions or, more generally,
of these invariant subspaces. If one of them turns out to be stable,
then its qualitative features are present for
a non-zero measure neighbourhood of initial conditions in phase space,
thus acquiring physical significance.

 From another point of view, it is reasonable to expect some relation
between loss of stability of periodic orbits (or more generally
of invariant subspaces) and the onset of chaos in the system.
This is suggested by the generic behaviour of near-integrable Hamiltonian
systems where, as the perturbation parameter is increased,
the ``chaotic sea'' expands in phase space, forcing regular regions to shrink.
Budinsky and Bountis~\cite{Budinsky},
conjectured that the destabilization of known periodic orbits
of an Hamiltonian system, although being only a local property,
could give an upper bound estimate for the onset of large-scale chaos.
This suggestion was perhaps not fully justified since, at that time,
only few one-mode solutions (e.g. $e=N/2$) were known, and moreover the
behaviour of very low-dimensional systems (e.g. H\'{e}non-Heiles)
had been studied.
It assumes more relevance now, since we know that several invariant
manifolds exists. However, the fact that periodic and quasiperiodic
solutions lose stability or invariant submanifolds become unstable does not
necessarily imply large-scale chaos. We will show that this loss of stability
leads to the formation of stochastic layers, but their thickness and extension
in phase space remain small at small $\epsilon$. In modal space, the
instability
of all these newly discovered submanifolds does not lead to energy
equipartition
if $\epsilon$ is sufficiently small, but to exponentially localized mode
energy spectra with irregular, chaotic, time dependencies of mode energies.

\subsection{Bifurcation and transition to chaos in a reduced system}
\label{N/4,N/2}

It is easy to verify that the set ${\cal F} = \{ N/4,N/2,3N/4 \}$
is of type I. Again, one has to verify condition~(\ref{type-I}) for
${\cal A}={\cal F}$. Apart from permutations of the indices, there are ten
distinct kinds of coefficients to consider, corresponding to the number
of triplets that can be formed with the elements of $\cal F$ without
regard to order and with allowed repetition of any element.
Five of them belong
to cases already treated in Secs.~\ref{One-mode} and~\ref{Two-mode}.
Examining also the remaining cases, we find that the only non-zero
coefficients having three indices in $\cal F$ and the remaining index
{\em unrestricted} (in $\cal M$) are, apart from permutations,
$C_{N/4,\,N/4,\,N/2,\,N/2}=C_{3N/4,\,3N/4,\,N/2,\,N/2}=2$
and, of course, the three $C_{eeee}$ with $e \in \cal F$.
This allows to conclude that $\cal F$ is of type I, and at the same time
leads to write the corresponding reduced Hamiltonian:
\bea
H_{\cal F} & = & \frac{1}{2}
\left( P_{N/4}^2 + P_{N/2}^2 + P_{3N/4}^2 \right) +
 Q_{N/4}^2  + 2Q_{N/2}^2 + Q_{3N/4}^2 +   \nonumber \\
 & &  \frac{2 \beta}{N} \left(Q_{N/4}^4  + 2Q_{N/2}^4 +
Q_{3N/4}^4+6Q_{N/4}^2 Q_{N/2}^2 + 6Q_{3N/4}^2 Q_{N/2}^2 \right).
\label{H-three-mode}
\eea
Either from the corresponding equations of motion, or from what has been
found above concerning the coefficients, we observe
that there exist also two new two-mode
invariant subspaces: that of the pair $\{N/4,N/2\}$
and that of $\{3N/4,N/2\}$.
At variance with cases discussed in Secs.~\ref{One-mode}--\ref{Two-mode},
all these new invariant sets give rise to
seemingly non-integrable Hamiltonians.
The case of a two-mode invariant subset is particularly interesting
because, for such a reduced system, the qualitative features of
the dynamics can be visualized on a two-dimensional Poincar\'e section.

The equations of motion corresponding to the set $\{N/4,N/2\}$ are:
\beq
\left\{  \begin{array}{lll}
\ddot{Q}_{N/4} & = &
{\displaystyle  -2Q_{N/4}-8 \frac{\beta}{N}Q_{N/4}^3-
24\frac{\beta}{N} Q_{N/4} Q_{N/2}^2  } \\
 & & \\
\ddot{Q}_{N/2} & = &
{\displaystyle  -4Q_{N/2}-16 \frac{\beta}{N}Q_{N/2}^3-
24\frac{\beta}{N} Q_{N/2} Q_{N/4}^2}~.
\end{array}
\right.
\label{N4N2}
\eeq
Those relative to the other set $\{3N/4,N/2\}$ are identical, substituting
$Q_{N/4} \rightarrow Q_{3N/4}$.
Rescaling the coordinates by $\sqrt{\beta/N}$ one obtains
equations independent of the ratio $\beta/N$.
Thus an orbit of the system~(\ref{N4N2}) is mapped to one of
the same system with $\beta/N=1$.
This means that the system~(\ref{N4N2}) remains
meaningful in the termodynamic limit.

We want, in particular, to investigate the stability of the
``out-of-phase'' nonlinear oscillation~(\ref{N/2old}), i.e. the
one-mode periodic solution~(\ref{e}) having $e=N/2$.
Therefore we choose the $(Q_{N/4},P_{N/4})$ plane
as the surface of section; in fact on this plane that solution
is represented by a fixed point for the Poincar\'e first return map,
located at $(Q_{N/4},P_{N/4})=(0,0)$.
The one-mode solution with $e=N/4$ lies, instead, on the surface of section and
can be drawn simply as the curve bounding the energetically accessible region.
In Figs.~1, 2 and~3 we show numerically generated Poincar\'e sections
for the system~(\ref{N4N2}), corresponding to three different increasing
values of the control parameter $\epsilon$ in~(\ref{epsilon}).
For each $\epsilon$ value, the intersections originated by several different
initial conditions are shown, together with the bounding curve corresponding
to mode $e=N/4$.

We observe that the fixed point at the origin is stable at
low values of $\epsilon$. When the energy density is increased,
a bifurcation occurs, giving rise to a pair of stable fixed points,
while the original one loses stability.
A small closed invariant curve which surrounds only one of them is plotted
in Fig.~2 to make clear the presence of two distinct
fixed points (i.e. two different bifurcated orbits for the flow),
so that the bifurcation is not period-doubling.
To confirm this we have also numerically computed
the eigenvalues $(\Lambda_{1},\Lambda_{2})$ of the Poincar\'e map linearized
around the origin, obtaining a more precise estimation of the critical
$\epsilon$ value.
As $\epsilon$ is increased the eigenvalues,
lying on the unit circle in the complex plane, move on it
until they collide at $\Lambda_{1}=\Lambda_{2}=+1$
for $\epsilon = 1.127$ and then move apart on the real axis.
After the fixed point at the origin turns unstable, a stochastic
layer is originated by homoclinic intersections,
too thin to be seen in Fig.~2 but already very thick in Fig.~3.
Therefore when the stationary wave~(\ref{N/2old}) has high enough amplitude
and is perturbed only by a small component along mode $N/4$,
the interaction between the two modes leads to chaotic behaviour.
However even at the $\epsilon$ value of Fig.~3 we do not observe transport
over the full phase space which is signature of large-scale chaos
(which, of course, will set in at larger values of $\epsilon$).
This shows that the loss of stability
of periodic orbits induces the appearence of stochastic layers but the
large-scale chaos phenomenon is not a direct consequence of it,
contrary to the expectations of Ref.~\cite{Budinsky}.

In the next Subsection we study the stability problem for the $e=N/2$ mode
in the full phase space.

\subsection{Large $N$ stability}
\label{Large-N-stability}

The problem of stability of a known solution $\overline{\Q}(t)$
under generic small perturbations in the initial data,
is tackled by studying the variational equations,
obtained by linearizing system~(\ref{eqmoto}) around $\overline{\Q}(t)$.
Let $\mbox{\boldmath $x$}(t)=\Q(t)-\overline{\Q}(t)$ be the
small separation vector between a solution $\Q(t)$, initially close to
$\overline{\Q}$, and the reference solution $\overline{\Q}(t)$ itself.
Then \mbox{\boldmath $x$} obey, in linear approximation, the following
system of equations:
\beq
\ddot{x}_r = \sum_{s=1}^{N-1} B_{rs}(t) x_s,~~~~~~r=1,\ldots,N-1~,
\label{eqlinearized}
\eeq
where
\beq B_{rs}(t)= \left. \frac{\partial F_r}{\partial Q_s}
\right|_{\Q=\overline{\Q}(t)}.
\eeq
is the Jacobian matrix of {\boldmath $F$} evaluated on the reference solution.
We limit ourselves to study here the stability of the periodic
single-mode $\{e\}$ solutions $\overline{Q}_j(t)=\delta_{je} Q_e(t)$,
with $Q_e(t)$ given by~(\ref{e}).
In this case system~(\ref{eqlinearized}) reduce to
\beq
\ddot{x}_r=-\omega^{2}_{r} x_r - \frac{3\beta}{2N} \omega_r \omega_{e}^{2}
Q_{e}^{2}(t) \sum_{s=1}^{N-1}
\omega_s C_{rs}^{(e)} \, x_s~,~~~~~~r=1,\ldots,N-1
\label{lin-around-e}
\eeq
where the coupling matrix, which depends on the chosen mode
$e$ in~(\ref{solitari}), has elements $C_{rs}^{(e)}\equiv C_{rsee}$.
It turns out that $C_{rs}^{(e)}$ is diagonal only for $e=N/2$, with
$C_{rs}^{(N/2)}=2\delta_{rs}$. The study of the stability of the
$e=N/2$ nonlinear mode is therefore simpler, because the different
components of the perturbation in modal space are all decoupled and
can be studied separately, each one being a problem of parametric
excitation described by an equation of the Hill type.
This permits approximate semi-analytic estimates also
for what concerns the dependence of stability properties on the
number $N$ of oscillators. Accordingly, we restrict here to the case $e=N/2$.
The study of other one-mode
cases (e.g. $e=N/3$) would imply the study of the stability for
a system of coupled linear equations with periodic coefficients,
a more complex problem of many degrees of freedom parametric excitation.
This could be done by producing numerically the appropriate transfer
matrix, but such a procedure should be repeated for each
fixed value of $N$ and the large $N$ result should be extrapolated.

For $e=N/2$, after using~(\ref{e}), Eqs.~(\ref{lin-around-e})
reduce to a set of decoupled Lam\'{e} equations~\cite{Whittaker,Arscott}
\beq
\ddot{x}_r = -\omega_r^2 \left[ 1+ 12 \frac{\beta A^2}{N}
\cn^2 (\Omega_{N/2} t, k) \right] x_r~,~~~~~~r=1,\ldots,N-1~,
\label{Lame-cn}
\eeq
with $\Omega_{N/2}$ and $k$ depending on $A$ as given in~(\ref{Omega})
and~(\ref{modulus}). The physical control parameter, proportional
to the energy density, is $\epsilon$, which is in a one-to-one relation
with $A$. However the analysis is simpler if we choose the modulus $k$ as
a control parameter. In fact, as $\epsilon$ is varied from zero to $+\infty$,
the modulus $k$ increases monotonically in the interval $[0,1/\sqrt{2}\,)$,
the limit $\epsilon\rightarrow +\infty$ corresponding
to $k \rightarrow 1/\sqrt{2}$. The relation $\epsilon(k)$ is
\beq
\epsilon=\frac{k^2}{1-2k^2} \left(1+\frac{k^2}{1-2k^2}\right)~.
\label{epsilon(k)}
\eeq
Expressing $\Omega_{N/2}$ and $\beta A^2/N$ as functions of $k$, and
after rescaling time at each fixed $k$ (which does not alter stability
properties), each of the equations in~(\ref{Lame-cn}) can be put in the
standard Jacobian form of Lam\'{e} equation~\cite{Whittaker,Arscott}
\beq
y''+[\alpha-\nu(\nu+1)k^2 \sn^2(u,k)]\,y=0~,
\label{Lame-standard}
\eeq
where the prime superscript denotes differentiation with respect to the
new time $u=\Omega_{N/2} t$ and $y$ stands for the generic variable $x_r$.
For the perturbation component $x_r$ in the direction of mode number $r$,
the parameters in~(\ref{Lame-standard}) are
\beq
\alpha=(1+4k^2)\omega_{r}^{2}/4~;~~~\nu(\nu+1)=3\omega_{r}^{2}/2~.
\label{alpha-omega}
\eeq
We observe that dependence on the mode number $r$ and on the number $N$
of oscillators comes only through the ratio
$r/N$ appearing in $\omega_r$ (see eq.~(\ref{omega})). This allows to
discuss in an unified way also the dependence on the system size $N$.
In fact, we can fix $\epsilon$ and then consider
two systems of different sizes $N_1$ and $N_2$; the linearized equations for a
mode $r_1$ in system $N_1$ and for a mode $r_2$ in system $N_2$
are the same if $r_1/N_1=r_2/N_2$.

Resuming, we discuss the stability of the zero solution
of~(\ref{Lame-standard})
as a function of two parameters $(\rho,k)$ with
$0\leq \rho<1$, $0\leq k<1/\sqrt{2}$, when $(\alpha,\nu)$ are given by
\beq
\alpha=\rho(1+4k^2)~;~~~\nu(\nu+1)=6\rho~.
\eeq
A given pair $(\rho,k)$ describes at the same time different mode numbers $r$
in systems with different $N$, all having the same $r/N$ value given
by the relation $\rho=\sin^2(\pi r/N)$, with the same $\epsilon$ for
the unperturbed solution $e=N/2$. We first consider $\rho$ as a continuous
parameter, which corresponds to the limit $N\rightarrow\infty$, and then
we discuss what happens when $\rho$ can take only a finite set of values,
as it is for any finite $N$.

For integer $\nu$ many rigorous results are known for
eq.~(\ref{Lame-standard})~\cite{Whittaker,Arscott}, and it is in particular
known~\cite{Magnus} that there are only $\nu+1$ instability regions
in the $(\alpha,k)$ plane.
However, in our case only the integer
values $\nu=1$ and $\nu=2$ are possible,
corresponding to $\rho=1/3$ and $\rho=1$, respectively.
The stability charts in the $(\alpha,k)$ plane for these two
integer $\nu$ cases of eq.~(\ref{Lame-standard}) have been explicitly
constructed in~\cite{Pecelli}.
The second case $\rho=1$, is of limited interest
to us since it corresponds to $r=N/2$, i.e. to perturbations along the
invariant subspace $e=N/2$; the results in~\cite{Pecelli} just
confirm that there exists a periodic solution for $x_{N/2}$ at all $k$ values,
as it must a priori be, since it corresponds to a perturbation along
the periodic orbit $Q_{N/2}(t)$ under study. More interesting is the
case $\rho=1/3$: the representative point in the $(\alpha,k)$ parameter
plane of~(\ref{Lame-standard}) with $\nu=1$
moves on the curve $\alpha=(1+4k^2)/3$
as $k$ is varied, and we find (see Fig.~4 of Ref.~\cite{Pecelli}) that it
lies always in the first stable region for $0\leq k<1/\sqrt{2}$.
It touches the boundary curve of the instability tongue only for $k=1/2$.
Therefore the mode corresponding to $\rho=1/3$ is stable
(as a perturbation to $e=N/2$) at
any finite energy density $\epsilon$, and reaches the stability
border only asymptotically for $\epsilon \rightarrow \infty$.

To solve the stability problem in the generic $\nu$ case we
approximately reduce eq.~(\ref{Lame-standard})
to Mathieu equation, whose stability
diagram is well known~\cite{Abramowitz,Arscott}.
In fact, in the parameter region
of our interest $0 < k^2 < 1/2$, the function
$\sn^2$ appearing in~(\ref{Lame-standard}) is well approximated by
its first order Fourier development. This latter is given by
\beq
k^2 \mbox{sn}^2(u,k) = 1 - \frac{E}{K}
-\left(\frac{\pi}{K}\right)^2 \sum_{n=1}^{\infty}
\frac{n}{\sinh(n \pi K'/ K)} \cos\left(\frac{n\pi u}{K}\right)~,
\label{sn2-Fourier}
\eeq
where $K$ and $E$ are the complete elliptic integrals
of first and second kind, respectively, with modulus $k$, while
$K'(k) = K (\sqrt{1-k^2}\,)$. The Fourier coefficients are exponentially
decreasing with order $n$, the more rapidly the smaller $k$ is.
In the worst case to consider, i.e. $k\rightarrow 1/\sqrt{2}$ the ratio
between the $n=2$ and $n=1$ coefficients is $\approx 0.086$.
Furthermore, we are interested in the modes which first go unstable, i.e.
which possibly destabilize at small $k$, and for them the error is much
smaller. Therefore, we truncate~(\ref{sn2-Fourier}) keeping only the $n=1$ term
and obtain, after a further time rescaling $\tau=\pi u/2 K$, the
canonical form of Mathieu equation:
\beq
\frac{d^2 y}{d\tau^2}+[a-2q\cos(2\tau)]\,y=0~,
\label{Mathieu}
\eeq
where
\beq
a=\rho \left(\frac{2K}{\pi}\right)^2  \left(-5+4k^2+6\frac{E}{K}\right)~;~~~
q=-\frac{12 \rho}{\sinh(\pi K'/K)}~.
\label{parametric-curves}
\eeq
Both parameters $a$ and $q$ depend on $k^2$ and $\rho$,
therefore for each fixed $\rho$ (which corresponds to fix a ratio $r/N$),
changing $k^2$ (and correspondingly the energy density $\epsilon$)
amounts to trace a curve in the $(q,a)$ parameter plane.
A stability transition happens if  this curve intersects characteristic
curves separating stable from unstable regions for the Mathieu equation.
In our case intersections can occur only with the characteristic curve
$b_1(q)$ (we use the standard notation, see e.g.~\cite{Abramowitz})
bounding from below the first tongue of instability.
In Fig.~4 we show the $(q,a)$ curves~(\ref{parametric-curves})
at fixed $\rho$ and varying $k$ for positive\footnote{We used positive
$q$ values for the
curves~(\protect\ref{parametric-curves}) since the Mathieu stability chart is
symmetric about the $a$-axis.}
$q$, and those at fixed $k$ and varying $\rho$,
together with the characteristic curves $b_1(q)$ and $a_1(q)$ bounding
the first instability tongue.
This mesh provides a reference frame to locate the values of
$\rho$ and $k$ corresponding to stable or unstable cases.
We do not trace the $\rho=1$ curve which superposes
with $a_1(q)$, showing only a slight deviation for $k^2>0.4$.
This superposition
is expected because the truncation to the first Fourier coefficients
in~(\ref{sn2-Fourier}) only slightly affect the exact solution of the Lam\'{e}
equation in the $\nu=1$ case.
We see that curves with fixed $\rho$ enter the instability region, for some
critical value $k_c(\rho)$ of $k$, only for $\rho$ larger than a value
such that the corresponding curve touches $b_1(q)$ only asymptotically
for $k\rightarrow \infty$. We numerically find such a value to be
$\rho \simeq 0.33$, which therefore must be identified with the value
$\rho=1/3$ having rigorously this property for the exact Lam\'{e}
equation, as seen before. This confirms the accuracy of the Mathieu
approximation. Also, we find that the $\rho=1/2$ ($r=N/4$) curve intersects
$b_1$ for $\epsilon\simeq 1.13$, in excellent agreement with the value
$\epsilon=1.127$ for the instability threshold in the $r=N/4$ direction,
obtained from the Poincar\'{e} map in Subsection~\ref{N/4,N/2}.

We have computed, for several values of $\rho>1/3$,
the critical $k_c(\rho)$ value at which
the corresponding constant $\rho$ curve intersects the $b_1$ line.
Returning to the physical control parameter $\epsilon$
by~(\ref{epsilon(k)}) we finally construct the marginal stability
curve $\epsilon_c(\rho)$ in Fig.~5.
Now we remember that, for a given number $N$ of oscillators,
only the discrete set of $\rho$ values
\beq
\rho=\sin^2(\pi r/N)~,~~~~r=1,\ldots,N-1
\label{rho-values}
\eeq
must be considered.
 From Fig.~5 we see that for each mode having $\rho>1/3$ there is a threshold
value $\epsilon_c(\rho)\not=0$ for instability.
Above $\epsilon_c(\rho)$ the $e=N/2$ nonlinear mode develops an instability
causing growth of the mode corresponding to $\rho$ through parametric
resonance.
On the contrary, modes with $\rho<1/3$ (i.e. $r/N<0.196$) are always stable
in the linear approximation
for any energy density of mode $e=N/2$, so that (nongeneric) perturbations
of the latter involving only these modes never lead to instability.
These modes, as well as modes with $\rho>1/3$ when $\epsilon<\epsilon_c(\rho)$,
can grow only if they are triggered
by the interaction with other modes that are unstable.
This kind of interaction is neglected in the linear approximation around
the $e=N/2$ mode, and comes into play only at later times, when unstable
modes have grown and consequently the linearized theory is no longer valid.

We can now describe what happens for a given system of $N$ particles.
First of all, for $N\geq 4$ there are always modes with $\rho>1/3$
so that the $e=N/2$ mode can never be stable for all energy densities.
As $\epsilon_c$ is a decreasing function of $\rho$, the first modes
to go unstable when $\epsilon$ is increased from zero are always
$r=(N/2)-1$ and $r=(N/2)+1$, which have $\rho=\cos^2(\pi/N)$.
Therefore for each (even) number $N$ of particles there is a non-zero
value $\tilde{\epsilon}(N)=\epsilon_c(\cos^2(\pi/N))$ below
which the nonlinear mode $e=N/2$ is stable. In fact for
$\epsilon<\tilde{\epsilon}(N)$ all the discrete points $(\rho,\epsilon)$
having abscissas~(\ref{rho-values}) are below the critical curve
$\epsilon_c(\rho)$ in Fig.~5.
Since $\epsilon_c\rightarrow 0$ for $\rho\rightarrow 1$, the critical
threshold  $\tilde{\epsilon}(N)$ for the stability of the nonlinear
out-of-phase mode approaches zero as the number $N$ of particles is
increased without limit.
Using power series expansions of the curve in~(\ref{parametric-curves})
and of $b_1(q)$ around the point $(q,a)=(0,1)$, we find that
$\tilde{\epsilon} (N) = \pi^2 / (3 N^2) + {\cal O} (N^{-4})$.
This is in agreement with the result obtained in Ref.~\cite{Budinsky}
using infinite Hill determinants. However, we do not agree with
their results in Table I, which seem to indicate that low modes become
unstable, while we claim that all modes with $\rho \leq 1/3$ are
always stable.

Furthermore, as the $\rho$ values~(\ref{rho-values}) to be considered
are increasingly dense as $N$ is increased, for each $\epsilon$,
no matter how small, there is a value $N$ for the system size above which
the $e=N/2$ mode is unstable. In this sense, we can say that the
$e=N/2$ mode is unstable at any nonzero value of the energy density
in the thermodynamic limit. However for very small $\epsilon$ it mantains a
metastable character, with a non-negligible lifetime, because the
largest growth rate of unstable modes vanishes as $\epsilon\rightarrow 0$.
This can be understood from Fig.~4 since for $k\rightarrow 0$ the unstable
part of the constant $k$ curve tends to the edge between the $b_1$ and $a_1$
lines and thus corresponds to a smaller and smaller growth
rate\footnote{We point out that our previous scalings of the time variable
do not change the order of magnitude of time scales at small $\epsilon$
since $\tau \approx 2t$ for $\epsilon \rightarrow 0$.}.

To confirm our analysis we have numerically integrated the full nonlinear
FPU model directly in normal coordinate space (Eqs.~(\ref{eqmoto})).
We have overcome the difficulty of the huge number
of interaction terms by first calculating a table of nonzero interaction
coefficient, whose number is only $\sim N^3$, which is then stored in a single
vector variable with a suitable label ordering\footnote{We thank J. Laskar
for suggesting this trick.}.
We excite mode $e=N/2$ with an energy $E_{N/2}$ corresponding
to a chosen $\epsilon$
and give to all other modes an equally distributed small amount of energy,
e.g. such that $\sum_{j\not=N/2} E_j/E_{N/2}\approx 10^{-13}$.
Then we follow the time evolution of all the modes, observing in particular
their harmonic energies. The latter are the physically relevant observables,
and are observed to grow exponentially for unstable modes,
while the corresponding normal coordinates oscillate with exponentially
growing envelope.

First of all we have verified the scaling suggested by~(\ref{alpha-omega}).
To compare results with different $N$ we mantain fixed the
ratio $\sum_{j\not=N/2} E_j/E_{N/2}$, so that
changing the $N$ value at fixed $\epsilon$ let modes
with the same $r/N$ ratio start with the same normal coordinate amplitude.
In Fig.~6 we show the modal energy spectrum at fixed
time for various values of $N$, with the energy density associated to mode
$e=N/2$ fixed at $\epsilon=0.1$.
The superposition of different spectra proves that only
direct interaction between mode $e=N/2$ and other modes,
through parametric excitation, is important in a first time evolution,
when the only growing modes are those predicted to be unstable on the basis
of the Mathieu equation stability diagram.
Modes with $r/N<0.35$ are predicted to be stable for $\epsilon=0.1$ from
the curve $\epsilon_c(\rho)$ shown in Fig.~5, in agreement with Fig.~6.
We have observed that at later times stable modes begin to grow
due to secondary interaction with the unstable modes grown meanwhile.

To understand the $\epsilon\rightarrow 0$ limit for the infinite chain
we have numerically computed the behaviour of the fastest growth rate of the
instability when $\epsilon$ is decreased while $N$ is increased.
In fact, if $\epsilon$ is decreased at fixed $N$, when it passes
below $\tilde{\epsilon}(N)$ no instability is detected. Therefore to
measure instability growth rates as $\epsilon$ decreases,
we must progressively increase $N$.
We have detected the fastest growing mode, $\bar{k}$, at each value
of $\epsilon$, and measured its exponential growth rate $\lambda$
($E_{\bar{k}} \sim 10^{\lambda t}$).
The dependence of $\lambda$ on $\epsilon$, for small $\epsilon$,
can be numerically fitted with a phenomenological law
$\lambda=c_1 \epsilon^{d} + c_2$ with $d=1.0$ $c_1=1.1$ and
$|c_2|<6\times 10^{-4}$ which linearly
extrapolates to a value consistent with zero in the limit
$\epsilon \rightarrow 0$. We have therefore an interpretation of what might
happen in the thermodynamic limit: in this limit the mode $e=N/2$
is always unstable under generic small perturbations but the rate of
instability goes to zero (a sort of marginal case). It is interesting
to observe that interchanging of the two limits $\epsilon \rightarrow 0$
and the thermodynamic limit $N\rightarrow \infty$ gives different answers
for the stability of the $e=N/2$ mode.
Stability is obtained if $\epsilon \rightarrow 0$
is performed first, while performing first the $N\rightarrow \infty$
limit we obtain instability.

An interesting and open question is the fate of the instability, since
our linearized theory describes the short time scale only.
For large $N$, the energy density stability threshold $\tilde{\epsilon}(N)$
is very small, and the $e=N/2$ mode is unstable for all relevant
$\epsilon$ values. However one should still distinguish different regimes.
For $\epsilon$ values large with respect to those of the instability,
the system evolves towards
equipartition, as it is in general known from studies on the threshold
of strong stochasticity for the FPU-$\beta$ system, although very little
attention has been paid to high mode initial excitations,
apart from Ref.~\cite{Izrailev66}.
However, for small $\epsilon$,
numerical results show that the (slow) growth of unstable modes saturates
at later times leading to a characteristic asymptotic form of the
energy spectrum where modal energies perform characteristic oscillations
and recurrences (similar to the well known FPU recurrences present for long
wavelength initial excitations). Furthermore the spectrum decays,
roughly exponentially, for modes far from mode number $N/2$, and it might
be possible the emergence of some coherent structure.
In this connection we point out that the more unstable modes are those
around the initially excited mode number, a characteristic feature
of modulational instability, well studied in
continuum systems such as, e.g., water waves~\cite{Benjamin}.
In the context of discrete systems different from FPU models, namely
chains made of {\em harmonically} coupled particles subjected to a
nonlinear {\em on-site} potential, modulational instability  has been
proposed~\cite{Peyrard} as the first step towards the creation of states
with spatially localized energy.
The possible existence of spatially localized oscillations,
called {\em intrinsic localized modes}, in FPU-type chains has been the object
of much work, starting from~\cite{intrinsic}, generally unrelated with
studies on equipartition and energy thresholds in the same systems.
Let us finally mention that in a recent interesting paper
by Sandusky and Page~\cite{Sandusky},
a connection is investigated between the existence of intrinsic
localized modes and the instability of the ``out-of-phase'' mode
in several kinds of one-dimensional homogeneous chains with anharmonic
intersite coupling.

\section{Conclusions}
\label{Conclusions}

There has been recently a revival of interest in conservative coupled
oscillator systems (see e.g.~\cite{intrinsic,Aubry}). In this paper we have
concentrated
our attention on the prototype of these systems, the Fermi-Pasta-Ulam
oscillator chain~\cite{Fput}. Although this system has been studied for
a long time, still it continues to be the source of many interesting physical
problems. The first Section of this paper is devoted to a comprehensive
review of old and new results (a recent historical reconstruction can
be found in~\cite{Ford-report}). Then, we turn to linear mode space analysis.
This was first performed in the basic paper by Izrailev and
Chirikov~\cite{Izrailev66}, although the scope was there different.
We have here derived a rather compact expression for mode coupling coefficients
(see formulas~(\ref{delta}) and~(\ref{Cijkl})), which leads to an efficient
writing of the Hamiltonian in mode space (see~(\ref{H1fin}) and~(\ref{Hfin})).
After an ``intermezzo'' devoted to the $N=3$ system, where we rederive
a well known integrability result by a one line proof, we pass to the central
issue of the paper: the existence of low-dimensional exact soutions and
invariant submanifolds. The method for proving their existence relies on
a detailed analysis of mode interaction coefficients and of selection rules
which prevent direct interactions among particular subsets of modes.
We have thus found one-mode and two-mode integrable Hamiltonians
(the latter because separable or rotation invariant) corresponding to
new standing and travelling nonlinear wave solutions
[see~(\ref{N/2old}),(\ref{N/4old}),(\ref{N43N4}),(\ref{sovrapposiz}),
(\ref{g-travelling}),(\ref{N3gamma}),(\ref{tw2})] of the FPU system, which hold
true in the thermodynamic limit. Three-mode invariant submanifolds are moreover
present (see~(\ref{H-three-mode})), for which a Poincar\'{e} section study
of two-mode associated subspaces reveals a bifurcation as
the control parameter $\epsilon$, proportional to energy per oscillator,
is varied. This bifurcation is induced by parametric perturbation and
leads to the formation of stochastic layers, corresponding to intermode
energy transfer. This effect suggests a physical interpretation of these
bifurcations, which could be responsible for opening channels of energy
transfer among modes. It should however be reminded that all submanifolds
so far found correspond to high $k$ modes. The mechanisms controlling
energy transfer for low $k$ modes seem to be completely different
(see Refs.~\cite{PRK95,DeLuca95,Parisi,DLR95}).

Moreover, the opening of a transfer channel does not necessarily mean that
energy flow is completed until energy equipartition is reached;
this can happen only if large-scale chaos is present, leading to the
diffusion of the orbits over the whole phase space.

A complete stability analysis is possible for mode $e=N/2$, after
reducing variational equations~(\ref{lin-around-e}) to a single Lam\'{e}
equation~(\ref{Lame-standard}) and a proper interpretation of control
parameters related to mode number, system size $N$ and energy density.
For finite $N$ we prove the existence of a low energy range of stability.
However this range shrinks to zero as $N\rightarrow \infty$ but,
correspondingly, the rate of instability goes to zero as
$\epsilon\rightarrow 0$, thus proving the presence of a sort of marginal
stability of mode $e=N/2$ in the thermodynamic limit at low energy density.
This is a hint to the belief that this solution might be physically relevant,
and moreover encourages to look for analogous stability proofs for other
one-mode and even two-mode solutions, whose existence was proven in this paper.

A more difficult matter would be the proof of stability of invariant
submanifolds. The odd and even invariant submanifolds mentioned at the
beginning
of Section~\ref{Explicit-solutions} also display an instability threshold.
This phenomenon has been studied in connection with an instability of
cnoidal waves in the modified KdV equation (mKdV) by Driscoll and
O'Neil~\cite{Driscoll}; the control parameter for this instability might be
different from energy density, as claimed in Ref.~\cite{DeLuca95}.
However the relation with this problem is not clear because the mKdV
describes only the long wavelength part of the mode spectrum.
The study of the stability of multi-mode invariant submanifolds
will be the subject of future investigations, together
with a more complete computer assisted search and classification of them.

\section*{Acknowledgments}

We acknowledge useful discussions with Jayme De Luca, Giovanni Gallavotti,
Allan Lichtenberg, Roberto Livi and Dima Shepelyansky.
We thank the Institute for Scientific
Interchange in Torino (Italy) for hospitality and use of computer facilities.
This work is part of the European Network on Stability and Universality
in Classical Mechanics (contract n. ERBCHRXCT940460).

\newpage

\section*{Figure captions}
\begin{description}

\item[Fig. 1.] Poincar\'{e} section for the system~(\protect\ref{N4N2})
on the $(Q_{N/4},P_{N/4})$ plane, with $\beta/N=0.4$. The energy value $E$
corresponds to $\epsilon=\beta E/N=1.0$. Each smooth curve corresponds
to a different initial condition on the constant energy shell.
The fixed point at the origin, representing the periodic
oscillation~(\protect\ref{N/2old}) on the lattice, is stable.

\item[Fig. 2.] Poincar\'{e} section for the system~(\protect\ref{N4N2})
on the $(Q_{N/4},P_{N/4})$ plane, with $\beta/N=0.4$. The energy value $E$
corresponds to $\epsilon=\beta E/N=1.25$.
The fixed point at the origin has become unstable and
a pair of new stable fixed points has bifurcated from it (see text
for details).

\item[Fig. 3.] Poincar\'{e} section for the system~(\protect\ref{N4N2})
on the $(Q_{N/4},P_{N/4})$ plane, with $\beta/N=0.4$. The energy value $E$
corresponds to $\epsilon=\beta E/N=10.0$.
The fixed point at the origin, which is by now unstable, lies within
a stochastic layer.

\item[Fig. 4.] Stability chart of the Mathieu equation~(\protect\ref{Mathieu}).
The instability region lies between the continuous lines. $b_1(q)$ is the
lower bound and $a_1(q)$ the upper one. The lines match at $(0,1)$.
We draw in the figure also the $(q,a)$ lines in
formula~(\protect\ref{parametric-curves}),
parametrized by $\rho$ and $k$. The
dashed lines correspond to a fixed $\rho$ and a continuosly varying $k$
 from 0 to $1/\sqrt{2}$; they are traced at $\rho$ intervals of $0.1$ from
$\rho=0$ to $\rho=0.9$. The dotted lines correspond instead to a fixed $k$
and a continuously varying $\rho$ from zero to one; the curves are traced
at $k^2$ intervals of $0.05$ from $0.05$ to $0.5$. Crossings of the lines
in formula~(\protect\ref{parametric-curves})
with the $b_1(q)$ line correspond to transitions to instability.

\item[Fig. 5.] Marginal stability curve $\epsilon_{c}(\rho)$ (full line).
The vertical asymptote at $\rho=1/3$ (dotted) is also traced to mark the
fact that all modes with $\rho<1/3$ are stable for any $\epsilon$ value.

\item[Fig. 6.] Modal linear energies $E_r$ vs. the ratio of the
mode number $r$ to the system size $N$ at fixed time $t=30$.
Different symbols correspond to different system sizes:
($+$) $N=8$; ($\times$) $N=16$; ($\Box$) $N=32$; ($\Diamond$) $N=64$.
The continuous curve is an interpolation to guide the eyes.
The initial values of modal energies are $3.2 \times 10^{-13}$.

\end{description}


\begin{thebibliography}{99}

\bibitem{Fput}
E. Fermi, J. Pasta, and S. Ulam, Los Alamos Report LA-1940 (1955),
later published
in {\it Collected Papers of Enrico
Fermi}, E. Segr\'e ed., University of Chicago Press, Chicago
(1965) (Vol. II, p. 978); also reprinted in
{\it Nonlinear Wave Motion}, A. C. Newell ed.,
Lect. Appl. Math. {\bf 15}, AMS, Providence, Rhode Island (1974);
also in
{\it The Many-Body Problem}, D. C. Mattis ed., World Scientific,
Singapore (1993).

\bibitem{Izrailev66}
F. M. Izrailev and B. V. Chirikov, Dokl. Akad. Nauk SSSR
{\bf 166} (1966) 57  [Sov. Phys. Dokl. {\bf 11} (1966) 30].

\bibitem{Chir-Izr-Tayursky73}
B. V. Chirikov, F. M. Izrailev, and V. A. Tayursky, Comp. Phys. Commun.
{\bf 5} (1973) 11.

\bibitem{KAM}
A. N. Kolmogorov, Dokl. Akad. Nauk SSSR {\bf 98} (1954) 527;
J. Moser, Nachr. Akad. Wiss. G\"ottingen Math. Phys. Kl.
{\bf 2} (1962) 1;
V. I. Arnol'd, Russ. Math. Surv. {\bf 18} (1963) 9 and 85.

\bibitem{Ben-Varenna}
G. Benettin, {\it Ordered and Chaotic Motions in Dynamical Systems
with Many Degrees of Freedom}, in {\it Molecular Dynamics Simulation
of Statistical Mechanical Systems}, Varenna XCVII Course,
G. Ciccotti and W. G. Hoover eds., North-Holland, Amsterdam (1986).

\bibitem{Livi85}
R. Livi, M. Pettini, S. Ruffo, M. Sparpaglione, and A. Vulpiani,
Phys. Rev. A {\bf 31} (1985) 1039;
R. Livi, M. Pettini, S. Ruffo, and A. Vulpiani, Phys. Rev.
A {\bf 31} (1985) 2740.

\bibitem{Kantz89}
H. Kantz, Physica D {\bf 39} (1989) 322.

\bibitem{Pettini90}
M. Pettini and M. Landolfi, Phys. Rev. A {\bf 41} (1990) 768.

\bibitem{Galgani92}
L. Galgani, A. Giorgilli, A. Martinoli, and S. Vanzini,
Physica D {\bf 59} (1992) 334.

\bibitem{Livi86}
R. Livi, A. Politi and S. Ruffo, J. Phys. A {\bf 19} (1986) 2033.

\bibitem{Eckmann}
J. P. Eckmann and E. Wayne, J. Stat. Phys. {\bf 50} (1988) 853.

\bibitem{Zabusky65}
N. J. Zabusky and M. D. Kruskal, Phys. Rev. Lett. {\bf 15} (1965) 240.

\bibitem{Gardner67}
C. S. Gardner, J. M. Greene, M. D. Kruskal, and R. M. Miura, Phys. Rev. Lett.
{\bf 19} (1967) 1095.

\bibitem{Lax68}
P. D. Lax, Comm. Pure Appl. Math. {\bf 21} (1968) 467.

\bibitem{Zakharov71}
V. E. Zakharov and L. D. Faddeev, Funct. Anal. Appl. {\bf 5} (1971) 280.

\bibitem{AblowitzKNS73}
M. J. Ablowitz, D. J. Kaup, A. C. Newell, and H. Segur, Phys. Rev. Lett.
{\bf 31} (1973) 125.

\bibitem{Ablowitz-books}
M. J. Ablowitz and H. Segur, {\it Solitons and the Inverse Scattering
Transform}, SIAM, Philadelphia (1981); M. J. Ablowitz and P. A. Clarkson,
{\it Solitons, Nonlinear Evolution Equations and Inverse Scattering},
Cambridge University Press (1991).

\bibitem{Ben-Ten}
G. Benettin, G. Lo Vecchio, and A. Tenenbaum, Phys. Rev. A {\bf 9}
(1979) 1252; G. Benettin, A. Tenenbaum, Phys. Rev. A {\bf 28} (1983) 3020.

\bibitem{CLMP95}
L. Casetti, R. Livi, A. Macchi, and M. Pettini,
{\it Relaxation times in an anharmonic crystal with diluted impurities},
submitted to Europhysics Letters (1994).

\bibitem{Tuck}
J. L. Tuck, Los Alamos Report  LA-3990 (1968);
J. L. Tuck and M. T. Menzel,
Adv. Math. {\bf 9} (1972) 399.

\bibitem{Ruelle82}
D. Ruelle, Commun. Math. Phys. {\bf 87} (1982) 287.

\bibitem{Manneville}
P. Manneville, in {\it Macroscopic Modeling of Turbulent Flows},
O. Pironneau ed., Lecture Notes in Physics {\bf 230}, 319, Springer-Verlag,
Berlin (1985).

\bibitem{Sinai}
Ya. G. Sinai, {\it A remark concerning the thermodynamical limit of
Lyapunov spectrum}, preprint (1995).

\bibitem{DeLuca95}
J. De Luca, A. J. Lichtenberg, and M. A.Lieberman, Chaos {\bf 5} (1995) 283.

\bibitem{KLR94}
H. Kantz, R. Livi and S. Ruffo, J. Stat. Phys. {\bf 76} (1994) 627.

\bibitem{DLR95} J. De Luca, A. J. Lichtenberg, and S. Ruffo,
Phys. Rev. E {\bf 51} (1995) 2877.

\bibitem{Pettini93}
M. Pettini, Phys. Rev. E {\bf 47} (1993) 838;
L. Casetti and M. Pettini, Phys. Rev. E {\bf 48} (1993) 4320.

\bibitem{Pettini95} L. Casetti, R. Livi, and M. Pettini, Phys. Rev. Lett.
{\bf 74} (1995) 375.

\bibitem{Nekho} N. N. Nekhoroshev, Funct. Anal. Appl. {\bf 5} (1971) 338;
		Russ. Math. Surv. {\bf 32} (1977) 1.

\bibitem{Parisi} F. Fucito, F. Marchesoni, E. Marinari, G. Parisi, L. Peliti,
S. Ruffo, and A. Vulpiani, J. de Physique {\bf 43} (1982) 707.

\bibitem{Livi83-exponential} R. Livi, M. Pettini, S. Ruffo, M. Sparpaglione,
and A. Vulpiani, Phys. Rev. A {\bf 28} (1983) 3544.

\bibitem{Shepelyansky}
D. L. Shepelyansky, unpublished (1994).

\bibitem{PRK95} P. Poggi, S. Ruffo, and H. Kantz,
Phys. Rev. E {\bf 52} (1995) 307.

\bibitem{Chirikov79}
B. V. Chirikov, Phys. Rep. {\bf 52} (1979) 263.

\bibitem{Ford} J. Ford, J. Math. Phys. {\bf 2} (1961) 387;
J. Ford and J. Waters, J. Math. Phys. {\bf 4} (1963) 1293;
E. A. Jackson, J. Math. Phys. {\bf 4} (1963) 551 and 686.

\bibitem{Sholl-Henry} D. S. Sholl and B. I. Henry, Phys. Lett. A {\bf 159}
(1991) 21.

\bibitem{Henry-Grindlay} B. I. Henry and J. Grindlay,
Physica  D {\bf 28} (1987) 49.

\bibitem{Saito} N. Ooyama, H. Hirooka, and N. Saito, J. Phys. Soc. Japan
{\bf 27} (1969) 815;
N. Saito, N. Ooyama, Y. Aizawa, and H. Hirooka,
Suppl. Prog. Theor. Phys. {\bf 45} (1970) 209;
N. Saito, N. Hirotomi, and A. Ichimura, J. Phys. Soc. Japan {\bf 39}
(1975) 1431.

\bibitem{Bivins} R. L. Bivins, N. Metropolis, and J. R. Pasta,
		  J. Comp. Phys. {\bf 12} (1973) 65.

\bibitem{Sholl} D. Sholl, Phys Lett. A {\bf 149} (1990) 253.

\bibitem{Sholl-tesi} D. S. Sholl, {\it Analytic Methods for Nonlinear
Lattices}, Bachelors thesis, The Australian National University (1991).

\bibitem{Chood}
G. V. Choodnovsky and D. V. Choodnovsky, Lettere al Nuovo Cimento
{\bf 19} (1977) 291.

\bibitem{Lawden} D. F. Lawden {\it Elliptic Functions and Applications},
Springer-Verlag, New York (1989).

\bibitem{Abramowitz} {\it Handbook of Mathematical Functions},
M. Abramowitz and I. A. Stegun eds., Dover, New York (1965).

\bibitem{Toda}
M. Toda, Phys. Rep. {\bf 18} (1975) 1.

\bibitem{Budinsky} N. Budinsky and T. Bountis,
Physica D {\bf 8} (1983) 445.

\bibitem{Whittaker} E. T. Whittaker and G. N. Watson,
{\it A Course of Modern Analysis}, 4th ed.,
Cambridge University Press, Cambridge (1946).

\bibitem{Arscott} F. M. Arscott, {\it Periodic Differential Equations},
Pergamon (1964).

\bibitem{Magnus} W. Magnus and S. Winkler {\it Hill's Equation},
Interscience, New York (1966).

\bibitem{Pecelli} G. Pecelli and E. S. Thomas,
Quart. Appl. Math. {\bf 36} (1978) 129.

\bibitem{Benjamin} T. B. Benjamin and J. E. Feir, J. Fluid Mech. {\bf 27}
(1967) 417.

\bibitem{Peyrard} Y. S. Kivshar and M. Peyrard, Phys. Rev. A {\bf 46} (1992)
3198; T. Dauxois and M. Peyrard, Phys. Rev. Lett. {\bf 70} (1993) 3935.

\bibitem{intrinsic} A. S. Dolgov, Sov. Phys. Solid State {\bf 28} (1986) 907;
A. J. Sievers and S. Takeno, Phys. Rev. Lett. {\bf 61} (1988) 970;
J. B. Page, Phys. Rev. B {\bf 41} (1990) 7835.

\bibitem{Sandusky} K. W. Sandusky and J. B. Page, Phys. Rev. B {\bf 50}
(1994) 866.

\bibitem{Aubry} R. S. MacKay, S. Aubry, Nonlinearity {\bf 7} (1994) 1623.

\bibitem{Ford-report} J. Ford, Phys. Rep. {\bf 213} (1992) 271.

\bibitem{Driscoll} C. F. Driscoll and T. M. O'Neil,
Phys. Rev. Lett. {\bf 37} (1976) 69;
Rocky Mountain J. of Math. {\bf 8} (1978) 211.

\end{thebibliography}
\end{document}